\documentclass[aps,prb,twocolumn,superscriptaddress,preprintnumbers, amsmath,amssymb,floatfix]{revtex4-2}

\usepackage{graphicx,color,amsmath,multirow,epstopdf,longtable}
\usepackage{array}

\DeclareGraphicsExtensions{.pdf,.eps,.png,.jpg,.mps}
\definecolor{yellow}{rgb}{1.,0.62,0.0}

\begin{document}

\title{Charge-carrier complexes in monolayer semiconductors}

\author{E.\ Mostaani}

\affiliation{Cambridge Graphene Centre, University of Cambridge, 9 J.\ J.\ Thomson Avenue, Cambridge, CB3 0FA, UK}

\author{R.\ J.\ Hunt}

\affiliation{Department of Physics, Lancaster University, Lancaster, LA1 4YB, UK}

\affiliation{Department of Engineering, Lancaster University, Lancaster, LA1 4YB, UK}

\author{D.\ M.\ Thomas}

\affiliation{Department of Physics, Lancaster University, Lancaster, LA1 4YB, UK}

\author{M.\ Szyniszewski}

\affiliation{Department of Physics, Lancaster University, Lancaster, LA1 4YB, UK}

\affiliation{Department of Physics and Astronomy, University College London, London, WC1E 6BT, UK}

\author{A.\ R. P.\ Montblanch}

\affiliation{Cavendish Laboratory, University of Cambridge, 19 J.\ J.\ Thomson Avenue, CB3 0HE, UK}

\author{M.\ Barbone}
\affiliation{Cambridge Graphene Centre, University of Cambridge, 9 J.\ J.\ Thomson Avenue, Cambridge, CB3 0FA, UK}

\affiliation{Cavendish Laboratory, University of Cambridge, 19 J.\ J.\ Thomson Avenue, CB3 0HE, UK}

\author{M.\ Atat\"ure}

\affiliation{Cavendish Laboratory, University of Cambridge, 19 J.\ J.\ Thomson Avenue, CB3 0HE, UK}

\author{N.\ D.\ Drummond}

\affiliation{Department of Physics, Lancaster University, Lancaster, LA1 4YB, UK}

\author{A.\ C.\ Ferrari}

\affiliation{Cambridge Graphene Centre, University of Cambridge, 9 J.\ J.\ Thomson Avenue, Cambridge, CB3 0FA, UK}

\begin{abstract}
The photoluminescence (PL) spectra of monolayer (1L) semiconductors feature peaks ascribed to different charge-carrier complexes. We perform diffusion quantum Monte Carlo simulations of the binding energies of these complexes and examine their response to electric and magnetic fields. We focus on quintons (charged biexcitons), since they are the largest free charge-carrier complexes in transition-metal dichalcogenides (TMDs). We examine the accuracy of the Rytova-Keldysh interaction potential between charges by comparing the binding energies of charge-carrier complexes in 1L-TMDs with results obtained using \textit{ab initio} interaction potentials. Magnetic fields$<8$T change the binding energies (BEs) by$\sim0.2$ meV\,T$^{-1}$, in agreement with experiments, with the BE variations of different complexes being very similar. Our results will help identify charge complexes in the PL spectra of 1L-semiconductors.
\end{abstract}
\maketitle
\section{Introduction}
The optical properties of layered semiconductors, such as transition-metal dichalcogenides (TMDs), change as the sample thickness is reduced from bulk (B) down to a single layer (1L)\cite{Ruppert_2014}. Indirect band gaps in B-TMDs are often observed to transition to direct band gaps in 1L\cite{Mak_2010}, accompanied by the emergence of photoluminescence (PL)\cite{Splendiani_2010}. Excitonic effects are enhanced in 1Ls relative to B-TMDs, due to reduction in electrostatic screening of the interactions between charge carriers\cite{Chernikov_2014}. In many 1L-semiconductors, including TMDs with honeycomb lattices, spin-orbit coupling splits the conduction (CB) and valence (VB) bands at their extrema at the K and K$'$ points of the Brillouin zone\cite{Kormanyos_2015}. This results in optically controllable spin and valley degrees of freedom\cite{Mak2012,Zeng2012,Xiaodong2014}. Valley polarization is retained for$>1$ns\cite{Mak2012}, ideal for quantum device applications, such as quantum light-emitting diodes\cite{Palacios2016,Palacios2017,Dang2020,Montblanch2021}. Localized single-photon emitters that can be controlled by electroluminescence\cite{Palacios2016,Montblanch2021} are also promising for quantum photonics.

The binding energy (BE) of an exciton may be calculated from first principles by solving the Bethe-Salpeter equation (BSE)\cite{Salpeter1951} on top of many-body perturbation theory calculations within the $GW$ approximation\cite{Ramasubramaniam_2012,Robert_2016,Thygesen_2017}, or by quantum Monte Carlo (QMC) methods\cite{Hunt_2018}. However, studying charge-carrier complexes, such as quintons, using these approaches is computationally expensive\cite{Needs_2020}. Instead, the effective-mass approximation\cite{stebe1998} can be used, whereby the ground-state energy is modelled by considering an electron (e) and a hole (h) interacting within a two-band model\cite{Xiao_2012}, and their effective masses are defined by experiment or by first principles band structure calculations. In effective-mass models of charge-carrier complexes in layered semiconductor materials (LSMs), it is crucial to take into account the two-dimensional (2d) nature of the electrostatic screening, as this modifies the form of the interaction between carriers\cite{Rytova_1965,Keldysh_1979}. The situation for LSMs differs from III-V semiconductor heterostructures with a thickness$>1\mu$m\cite{Harrison_1985}, in which the Coulomb $1/r$ interaction between charge carriers scales down with the permittivity of the host material\cite{Harrison_1985}. In LSMs, the so-called Rytova-Keldysh interaction (RKI) potential\cite{Rytova_1965,Keldysh_1979} provides a more accurate interaction between charge carriers. Refs.\onlinecite{Velizhanin_2015,Szyniszewski_2017, Mostaani_2017} studied the formation of multicarrier bound states in 1L-semiconductors using QMC methods, such as path integral Monte Carlo\cite{Ceperley1995} and diffusion Monte Carlo (DMC)\cite{Needs_2020} to solve the Schr\"{o}dinger equation for quasiparticles interacting via the RKI potential. DMC is particularly powerful in studies of complexes with distinguishable quasiparticles\cite{Kittel2004}, as it is numerically exact in this case\cite{Foulkes_2001}. Ref.\onlinecite{Mostaani_2017} used DMC to predict the stability of negative quintons in TMDs with distinguishable charge carriers (in which all three e species have different spin and/or valley degrees of freedom). These predictions were confirmed in experimental studies, which provided evidence of quintons in hBN-encapsulated 1L-WSe$_2$\cite{Barbone_2018,Ziliang_2018, Li_2018, Chen_2018}, 1L-MoSe$_2$ on sapphire\cite{Hao_2017}, and 1L-WSe$_2$ on Si/SiO$_2$\cite{LiT_2018}.

In Mo and W-TMDs, the VB spin-splitting is sufficiently large that the lower spin-split bands are always occupied at room temperature (RT)\cite{Kormanyos_2015}, while the CB spin-splitting is comparable with RT\cite{Kormanyos_2015}. As a result, there are effectively 4 e and 2 h species available to form charge-carrier complexes at and below RT\cite{Mostaani_2017}.

\begin{figure}
\centerline{\includegraphics[width=90mm]{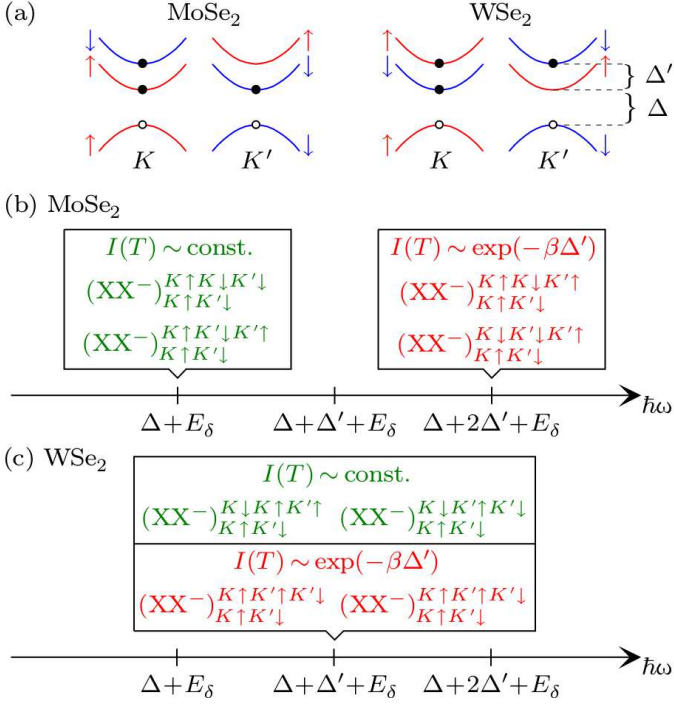}}
\caption{(a) Upper spin-split VB and spin-split CB for 1L-MoSe$_2$ and 1L-WSe$_2$. The spin-split VB is$>150$meV\cite{Kormanyos_2015}, so we only show the upper VB. (b,c) Classification of quinton recombination processes in 1L-Mo and 1L-W-TMDs. $E_\delta=E_{\rm XX^-}-E_{\rm X^-}$ is the difference between the total energies $E_{\rm XX^-}$ and $E_{\rm X^-}$ of XX$^-$ and X$^-$. $\hbar\omega$ indicates the photon energies at which XX$^-$ peaks in PL spectra are expected. ${({\rm XX}^-)}^{k_1\sigma_1k_2\sigma_2k_3\sigma_3}_{k_4\sigma_4k_5\sigma_5}$ denotes a quinton consisting of CB e in valleys $k_1$, $k_2$, and $k_3$ with spins $\sigma_1$, $\sigma_2$, $\sigma_3$ and VB h in valleys $k_4$ and $k_5$, with spins $\sigma_4$, $\sigma_5$. E.g., the quintons in (a) are both ${({\rm XX}^-)}_{K\uparrow K'\downarrow}^{K\uparrow K\downarrow K'\downarrow}$. Unlike Figs.1,2 of Ref.\onlinecite{Mostaani_2017}, we only show complexes with distinguishable charge carriers, because they are stable and should be experimentally observable.\label{fig:quint_class}}
\end{figure}

One can distinguish dark\cite{Efros_1996}, bright\cite{Efros_1996}, and semidark\cite{Danovich2017} charge-carrier complexes in TMDs. In dark complexes (Fig.\ref{fig:quint_class}a), radiative e-h recombination is not allowed due to spin and/or momentum mismatch between the constituent e/h\cite{Wang2017}, while in bright complexes (Fig.\ref{fig:quint_class}b), direct radiative e-h recombination is allowed by conservation of linear and angular momentum\cite{Qiu2013}. In semi-dark complexes (Fig.\ref{fig:quint_class}c), radiative recombination can take place following an intervalley scattering event assisted by a phonon that maintains spin, but swaps an e, e.g., from valley K$'$ to K\cite{Danovich2017}, accompanied by an energy shift due to the change in occupation of the upper and lower spin-split bands\cite{Danovich2017}.

Due to the nature of the CB spin-splitting of Mo-TMDs\cite{Kormanyos_2015} (Fig.\ref{fig:quint_class}a), bright states are energetically lower than dark\cite{Selig2018}. Hence, at low temperature $T<100$K, e in exciton (X), negative trion (X$^-$), and biexciton (XX) complexes occupy the lower spin-split bands. X complexes therefore travel only a small distance, e.g.$\sim1\mu$m in 1L-Mo-TMDs\cite{Hotta_2020,Uddin2020}, before radiative recombination, which reduces the chance to bind with another charge-carrier complex\cite{Robert_2018}. Furthermore, the XX PL peak may be difficult to distinguish from that of X$^-$, due to the small energy difference$\sim10$meV between their BEs\cite{Hotta_2020}. Ref.\onlinecite{Hao_2017} detected XX and quintons (XX$^-$) in 1L-MoSe$_2$ by 2d coherent spectroscopy (2dCS)\cite{Jonas_2003}. This method can focus on a delay time$\sim$10ps, over which XX or XX$^-$ are likely to form\cite{Bristow2009,Nardin2015}. However in 1L-W-TMDs, the most energetically stable excitonic states are dark\cite{Selig2018}, so that X have longer lifetimes ($\sim$1ps)\cite{Selig2018} than in 1L-Mo-TMDs ($\sim$0.5ps)\cite{Selig2018}, favouring larger than X charge-carrier complexes. We therefore focus on 1L-W-TMDs when comparing theory with experiments.

Fig.\ref{fig:quint_class} classifies ${\rm XX}^-$ in 1L-Mo- and W-TMDs with respect to recombination energy and $T$-dependence of the emitted photons' intensity. There are two XX$^-$ types: (1) those with 1e in the upper spin-split CB and 2e in the lower spin-split CB, and (2) those with 1e in the upper spin-split CB and 1e in the lower spin-split CB. The CB spin splittings in 1L-Mo- and W-TMDs are$\sim3$meV\cite{Kormanyos_2015} and$\sim30$meV\cite{Kormanyos_2015}, respectively. These are much less than the XX$^-$ BEs$\sim50$meV\cite{Hao_2017,Paur_2019,Barbone_2018,Paur_2019}, as reported in Table \ref{table:compare_quint}. The fact that the XX$^-$ BE is larger than the spin splitting implies XX$^-$ complexes are thermodynamically stable at T close to 0K, even taking into account the energy required to excite 1e to the upper spin-split CB. Assuming the CB spin-orbit splitting $\Delta'$ of 1L-TMDs to be$\ll$XX$^-$ BE, $E^\text{b}_{\text{XX}^-}$, each XX$^-$ can be treated as a two-state system\cite{Landau_1976}. For $k_\text{B}T\ll\Delta'\ll E^\text{b}_{\text{XX}^-}$, with $k_\text{B}$ the Boltzmann's constant, the fraction of XX$^-$ with 1e and 2e in the upper spin-split CB, hence the PL intensity of the corresponding XX$^-$, is\cite{Grahn1999}:
\begin{equation}
I(T)\sim \begin{cases} {\rm const.} & \text{for 1 e in upper spin band} \\ e^{-\Delta'/(k_{\rm B}T)} & \text{for 2 e in upper spin band}
\end{cases}
\label{eq:quinton_intensity}
\end{equation}

Here, we use DMC within the effective-mass approximation to calculate XX$^-$ energies in 1L-semiconductors. XX$^-$ are the largest free charge-carrier complexes in TMDs\cite{Mostaani_2017}. We provide an interpolation formula for XX$^-$ BEs for all 1L-semiconductors as a function of e and h effective masses, permittivity of the surrounding media, and in-plane susceptibility of the 1L-semiconductor. We also use DMC to calculate the energies of charge-carrier complexes in the presence of out-of-plane magnetic and in-plane electric fields, to identify whether the behavior in external fields can be used to investigate PL peaks. We find that applying an external magnetic field helps identifying charge-carrier complexes in 1L-semiconductors which have different e and h effective masses, while electric fields can be used to identify charge-carrier complexes in all 1L-semiconductors. We explore the accuracy of the RKI potential by comparing BEs with results obtained using \textit{ab initio} random-phase approximation (RPA) interaction potentials\cite{Aghajanian_2018}. We find that, within the effective-mass approximation, RKI can describe quasiparticles on length scales larger than the lattice constant. Therefore our results can be used to determine the PL spectra of excitonic charge complexes.
\begin{table*}
\caption{$E^{\rm b}_{{\rm XX}^-}$ of XX$^-$ calculated by DMC and by Eq.\ref{eq:quint_be}, compared with experiments from Refs.\onlinecite{Hao_2017,Paur_2019,LiT_2018,Barbone_2018}. The e and h effective masses $m_{\rm e}$ and $m_{\rm h}$ in terms of the free e mass $m_0$ are taken from many-body $GW$ calculations\cite{Ramasubramaniam_2012,Rasmussen2015}. We assume the materials suspended in vacuum, or encapsulated in hBN, or placed on top of a substrate such as SiO$_2$\cite{LiT_2018} and sapphire\cite{Hao_2017}. Wherever $\epsilon \neq \epsilon_0$, the vacuum $r_*$ is used}
\begin{center}
\begin{tabular}{lr@{}lr@{}lcr@{}lr@{}lr@{}lc}
\hline \hline
 & & & & & & & & \multicolumn{5}{c}{BE of XX$^-$ ($E_{\rm XX^-}^{\rm
    b}$) (meV)} \\
 \raisebox{1.5ex}[0pt]{TMD} &
 \multicolumn{2}{c}{\raisebox{1.5ex}[0pt]{$\frac{m_{\rm e}}{m_0}$}} &
 \multicolumn{2}{c}{\raisebox{1.5ex}[0pt]{$\frac{m_{\rm h}}{m_0}$}} &
 \raisebox{1.5ex}[0pt]{$\epsilon$} &
 \multicolumn{2}{c}{\raisebox{1.5ex}[0pt]{Vacuum $r_*$ ({\AA})}} &
 \multicolumn{2}{c}{DMC} &
 \multicolumn{2}{c}{Eq.\ (\ref{eq:quint_deex})} & Experiment \\
\hline
1L-MoS$_2$ (vac.) & $0$&$.35$ & $0$&$.428$ \cite{Cheiwchanchamnangij_2012} & $\epsilon_0$ & ~$38$&$.62$ \cite{Cheiwchanchamnangij_2012} & $58$&$.6(6)$ & ~~$59$&$.33$ & \\
\hline
1L-MoSe$_2$ (vac.) & $0$&$.38$ & $0$&$.44$ \cite{Shi_2013} & $\epsilon_0$& $39$&$.79$ \cite{Kumar_2012} & $57$&$.0(4)$ & $58$&$.13$&  \\
\hline
 & $0$&$.38$ & $0$&$.44$ \cite{Shi_2013} & $\epsilon_0$& $52$&$.2$ \footnote{The experimental XX and X$^-$ BEs are$\sim18$meV\cite{Hao_2017} and$\sim27$meV\cite{Hao_2017}, respectively. With $r_*=52.2$ {\AA} and $\epsilon=\epsilon_0$, Eqs.48,49 of Ref.\onlinecite{Mostaani_2017} give XX and X$^-$ BEs$\sim17.9$, 27.6meV, respectively.} & & & $46$&$.1$ & \\
\raisebox{1.5ex}[0pt]{1L-MoSe$_2$ (sapph.\ subs.)}& $0$&$.38$ & $0$&$.44$ \cite{Shi_2013} & $4.95\epsilon_0$& $39$&$.79$ \cite{Kumar_2012} & & & $29$&$.3$ & \raisebox{1.5ex}[0pt]{$\sim 40$ \cite{Hao_2017}}\\
\hline
1L-MoTe$_2$ (vac.) & $0$&$.65$ & $0$&$.64$ \cite{Rasmussen2015} & $\epsilon_0$ & $73$&$.61$ \cite{Ilkka_2015} & $33$&$.8(3)$ & $35$&$.61$ & \\
\hline
1L-WS$_2$ (vac.) & $0$&$.27$ & $0$&$.32$ \cite{Shi_2013} & $\epsilon_0$ & $37$&$.89$ \cite{Berkelbach_2013} & $57$&$.4(3)$ & $57$&$.60$ & \\
\hline
& $0$&$.27$ & $0$&$.32$ \cite{Shi_2013}& $\epsilon_0$ & $45$&$.1$ \footnote{The experimental XX and X$^-$ BEs are$\sim19.2$meV\cite{Paur_2019} and$\sim30.2$meV\cite{Paur_2019}, respectively. With $r_*=45.1${\AA} and $\epsilon=\epsilon_0$, Eqs.48,49 of Ref.\onlinecite{Mostaani_2017} give XX and X$^-$ BEs$\sim$19.9, 29.2meV, respectively.} & & & $49$&$.7$ & \\
\raisebox{1.5ex}[0pt]{1L-WS$_2$ (hBN)} & $0$&$.27$ & $0$&$.32$ \cite{Shi_2013} & $4\epsilon_0$ & $37$&$.89$ \cite{Berkelbach_2013} & & & $31$&$.4$ & \raisebox{1.5ex}[0pt]{$52.4$ \cite{Paur_2019}}\\
\hline
1L-WSe$_2$ (vac.) & $0$&$.29$ & $0$&$.34$ \cite{Shi_2013} & $\epsilon_0$& $45$&$.11$ \cite{Berkelbach_2013} & $52$&$.0(7)$ & $50$&$.23$ & \\
\hline
& $0$&$.29$ & $0$&$.34$ \cite{Shi_2013} &$\epsilon_0$& $44$&$.33$ \footnote{The experimental X$^-$ BE is$\sim30$meV\cite{LiT_2018}. With $r_*=44.33${\AA} and $\epsilon=\epsilon_0$, Eq.49 of Ref.\onlinecite{Mostaani_2017} gives BE$\sim30$meV.}  & & & $50$&$.97$ & \\
\raisebox{1.5ex}[0pt]{1L-WSe$_2$ (SiO$_2$ subs.)} & $0$&$.29$ & $0$&$.34$ \cite{Shi_2013} & $2.45\epsilon_0$& $45$&$.11$ \cite{Berkelbach_2013} & & & $37$&$.6$ & \raisebox{1.5ex}[0pt]{$51$ \cite{LiT_2018}}\\
\hline
& $0$&$.29$ & $0$&$.34$ \cite{Shi_2013} & $\epsilon_0$& $48$& \footnote{The experimental XX BEs are$\sim18.2$meV\cite{Barbone_2018} or$\sim20.1$meV\cite{Paur_2019}, and those of X$^-$ are$\sim27.1$meV\cite{Barbone_2018} or$\sim29.7$meV\cite{Paur_2019}. With $r_*=48${\AA} and $\epsilon=\epsilon_0$, Eqs.48,49 of Ref.\onlinecite{Mostaani_2017} give XX and X$^-$ BEs$\sim$18.9, 28.1meV, respectively.} & & & $47$&$.7$ & \\
\raisebox{1.5ex}[0pt]{1L-WSe$_2$ (hBN)} & $0$&$.29$ & $0$&$.34$ \cite{Shi_2013} & $4\epsilon_0$& $45$&$.11$ \cite{Berkelbach_2013}& & & $28$& & \raisebox{1.5ex}[0pt]{$49$ \cite{Barbone_2018}, $50.7$ \cite{Paur_2019}} \\
\hline
1L-WTe$_2$ (vac.) & $0$&$.325$ & $0$&$.460$ \cite{Kormanyos_2015} & $\epsilon_0$&$49$&$.56$ \cite{Kumar_2012,Lu_2016} & $47$&$.5(3)$ & $48$&$.56$ & \\
\hline \hline
\end{tabular}
\label{table:compare_quint}
\end{center}
\end{table*}
\begin{table}[!htbp]
\caption{List of acronyms}
\begin{center}
\renewcommand*{\arraystretch}{1.}
\begin{longtable}{ll}
\hline \hline
  Symbol  & Definition  \\
\hline \hline
\multicolumn{2}{c}{General acronyms} \\
\hline
2d  & two-dimensional \\
TMD & Transition metal dichalcogenide\\
1L & Monolayer \\
ML & Multilayer \\
LSMs & Layered semiconductor materials \\
RT & Room temperature \\
BE  & Binding energy \\
VB & Valence band \\
CB & Conduction band \\
PL & Photoluminescence \\
e.u. & Hartree excitonic units \\
SI   & International system\\
$\mu$  & Reduced mass \\
$m_{\rm e}$ & Electron effective mass \\
$m_{\rm h}$ & Hole effective mass \\
$m_0$ & Free electron mass \\
CoM  & Center of mass \\
$\epsilon$ & Absolute permittivity \\
$\hbar$ & Dirac constant \\
$r_\ast$  & Screening length \\
$B$  & Magnetic flux density \\
$F$  & Electric field \\
$R_y^\ast$ & Exciton Rydberg constant \\
$a_0^\ast$ & exciton Bohr radius \\
\hline
\multicolumn{2}{c}{Charge complexes} \\
\hline
e, h & Single electron, single hole \\
  X     & Exciton\\
  X$^-$ & Negative trion (negatively charged exciton)\\
  X$^+$ & Positive trion (positively charged exciton)\\
  XX    & Biexciton (bound states of 2e and 2h)\\
  XX$^-$ & Quinton (bound states of 3e and 2h)\\
  D$^-$XX & Donor-bound double-negative biexciton \\
  D$^0$X$^-$ & Donor-bound double-negative exciton \\
\hline
  \multicolumn{2}{c}{Total energies} \\
\hline
  E$_{\rm e}$ & Ground-state total energy of electron \\
  E$_{\rm h}$ & Ground-state total energy of hole \\
  E$_{\rm X}$ & Ground-state total energy of exciton \\
  E$_{\rm X^-}$ & Ground-state total energy of negative trion \\
  E$_{\rm XX}$  & Ground-state total energy of biexciton \\
  E$_{\rm XX^-}$ & Ground-state total energy of quinton \\
\hline
  \multicolumn{2}{c}{Binding energies} \\
  \hline
  E$_{\rm X}^{\rm b}$ & Binding energy of exciton \\
  E$_{\rm X^-}^{\rm b}$ & Binding energy of negative trion \\
  E$_{\rm XX}^{\rm b}$ & Binding energy of biexciton \\
  E$_{\rm XX^-}^{\rm b}$ & Binding energy of quinton \\
  E$_{\rm XX^-}^{\rm DE}$ & Deexcitonization energy of quinton \\
  E$_{\rm XX}^{\rm EA}$ & Electron affinity of biexciton \\
\hline
  \multicolumn{2}{c}{Methods} \\
  \hline
QMC & Quantum Monte Carlo  \\
DMC &  Diffusion Monte Carlo \\
VMC & Variational Monte Carlo \\
BSE & Bethe-Salpeter equation \\
RKI & Rytova-Keldysh interaction \\
RPA &  Random-phase approximation \\
RPAI & RPA interaction \\
FEM  & Finite-element method \\
2dCS & 2d coherent spectroscopy \\
CW  & Continuous wave \\
PVD & Physical vapor deposition \\
\hline \hline
\end{longtable}
%\end{tabular}
\label{tab:terms}
\end{center}
\end{table}
\section{Results and discussion \label{sec:results}}
\subsection{Units\label{sec:units}}
In the following, we will use Hartree excitonic units (e.u.), in which the e-h reduced mass $\mu$, $4\pi$ times the absolute permittivity $\epsilon$, the Dirac constant $\hbar$, and the charge $e$ are all equal to 1, i.e., $\mu=4\pi\epsilon=\hbar=e=1$. This helps to scale down the BEs with respect to $r_*$ and effective masses, as explained in Methods. The screening length is $r_* \equiv \kappa/(2\epsilon)$, with $\kappa$ the in-plane susceptibility, as discussed in Methods. The e.u. of length is the exciton Bohr radius $a_0^*=4\pi\epsilon\hbar^2/(\mu e^2)$\cite{Griffiths2016}, that of magnetic flux density is $B^*=\mu^2e^3/[{(4\pi\epsilon)}^2\hbar^3]$, that of electric field is $F^*=\mu^2e^5/[{(4\pi\epsilon)}^3\hbar^4]$, and that of energy is the exciton Hartree $2R_{\rm y}^*$, with $R_{\rm y}^*=\mu e^4/[2{(4\pi\epsilon)}^2\hbar^2]$ the exciton Rydberg constant\cite{Griffiths2016}.

For the logarithmic approximation to the RKI we will use a different set of units, as explained in Ref.\onlinecite{Mostaani_2017}. Since in the logarithmic regime $r\ll r_*$, where r is the separation between charge carriers, the behavior of the energy changes when compared with the intermediate regime $r\gg r_*$. In the logarithmic e.u., the e-h reduced mass $\mu$, $4\pi\epsilon r_*$, $\hbar$, and the electronic charge are all equal to 1, i.e., $\mu=4\pi\epsilon r_*=\hbar=e=1$. We define the logarithmic e.u. of length to be $\sqrt{2}r_0$, where $r_0=\sqrt{4\pi\epsilon r_*\hbar^2/(2e^2\mu)}$, the unit of energy $E_0=e^2/(4\pi\epsilon r_*)$, the unit of magnetic flux density $B_0=\sqrt{\mu E_0}/\left(\sqrt{2}er_0\right)=e\mu/(4\pi\epsilon r_*\hbar)$, and the unit of electric field $F_0=E_0/\left(\sqrt{2}r_0e\right)=\sqrt{e^4\mu/[{(4\pi\epsilon r_*)}^3 \hbar^2]}$. 

To convert the Bohr radius, flux density, electric field, and energy from e.u. to SI units, each value needs to be multiplied by $a_0^*$, $B^*$, $F^*$, and $2R_{\rm y}^*$, respectively. To convert from logarithmic e.u. to SI units, each value needs to be multiplied by $\sqrt{2}r_0$, $B_0$, $F_0$, and $E_0$, respectively.

Table II summarizes all acronyms used in this paper.
\subsection{Binding energies}
\label{sub_be_quint}
We define the X, X$^-$, and XX BEs as:
\begin{align}
E_{\rm X}^{\rm b}&=E_{\rm e}+E_{\rm h}-E_{\rm X} \\ E_{{\rm X}^-}^{\rm b}&=E_{\rm e}+E_{\rm X}-E_{{\rm X}^-} \label{eq:be} \\ E_{\rm XX}^{\rm b}&=2E_{\rm X}-E_{\rm XX}
\end{align}
where the complexes are defined in Table \ref{tab:terms}. In the absence of external fields, $E_{\rm  e}=E_{\rm h}=0$.

We define the de-excitonization energy of XX$^-$ as:
\begin{equation}
E_{{\rm XX}^-}^{\rm DE}=E_{\rm X}+E_{{\rm X}^-} -E_{{\rm XX}^-},
\label{eq:deex}
\end{equation}
and the electron affinity of XX as:
\begin{align}
E_{\rm XX}^{\rm EA}&=E_{\rm XX}+E_{\rm e}-E_{{\rm XX}^-}\nonumber
\\ &=E_{{\rm X}^-}^{\rm b}-E_{\rm XX}^{\rm b}+E_{{\rm XX}^-}^{\rm DE}.
\label{eq:af}
\end{align}
Since the most stable dissociated complexes have the lowest ground-state energies, the XX$^-$ BE is the minimum of $E_{{\rm XX}^-}^{\rm DE}$ and $E_{\rm XX}^{\rm EA}$ for a given $r_*$ and effective mass:
\begin{equation}
E_{{\rm XX}^-}^{\rm b}=\min\left\{E_{\rm XX}^{\rm EA},E_{{\rm XX}^-}^{\rm DE}\right\}.
\label{eq:quint_be}
\end{equation}
By comparing Eq.\ref{eq:quint_be} with \ref{eq:deex},\ref{eq:af} for 1L-TMDs, the energy difference between bright X and XX$^-$ PL peaks is $E_{{\rm XX}^-}^{\rm DE}$.

We calculate $E_{{\rm XX}^-}^{\rm DE}$ of XX$^-$ complexes in all 1L-semiconductors with all the possible values for $r_*/a_0^*=\{0,0.5,1,2,4,6,8,\infty\}$ and $\sigma=\{0, 0.1,0.2,\ldots, 1, 1.5, 4, 9, \infty\}$, where $\sigma=m_{\rm e}/m_{\rm h}$ is the mass ratio. We fit:
\begin{equation}
\frac{E_{\rm XX^-}^{\rm DE}}{R_{\rm y}^*(1-y)}=\frac{\sum\limits_{i=0}^4\sum\limits_{j=0}^{5-i} a_{ij}x^{i}y^j +b_1\sqrt x+b_2\sqrt{1-x}} {1+\sum\limits_{k=1}^3c_k y^k + y^2 \left(d_1\sqrt x+d_2 \sqrt{1-x}\right)}
\label{eq:quint_deex}
\end{equation}
to our DMC $E_{{\rm XX}^-}^{\rm DE}$, where $\{a_{ij}\}$, $\{b_i\}$, $\{c_i\}$, and $\{d_i\}$ are fitting parameters, $x=\sigma/(\sigma+1)=m_{\rm e}/\left(m_{\rm e}+m_{\rm h}\right)$ is a rescaled mass ratio and $y=r_*/(r_*+a_0^*)$ is a rescaled in-plane susceptibility parameter. The fitting function goes as the square root of the mass at extreme mass ratios ($\sigma=0$ and $\sigma=\infty$), as required by the Born-Oppenheimer approximation\cite{Griffiths2016}. We fit $E_{\rm XX^-}^{\rm DE}/[R_{\rm y}^*(1-y)]$ so that the asymptotic behavior at $r_*\to\infty$ obtained using the logarithmic interaction can be included in the fit. The error in the fitted $E_{{\rm XX}^-}^{\rm DE}$ is$<5$\% at each data point. The statistical error bars on the DMC $E_{{\rm XX}^-}^{\rm DE}$ data points are much smaller than the error in the fit. We therefore use an unweighted least-squares fit\cite{Krijnen1996}. We provide a program \cite{mostaani2022} that can be used to evaluate $E_{{\rm XX}^-}^{\rm DE}$ and XX$^-$ BE for any 1L-semiconductor, for which effective masses, $r_*$, and dielectric constant of the environment are the inputs.

\begin{figure}
\includegraphics[clip,width=87mm]{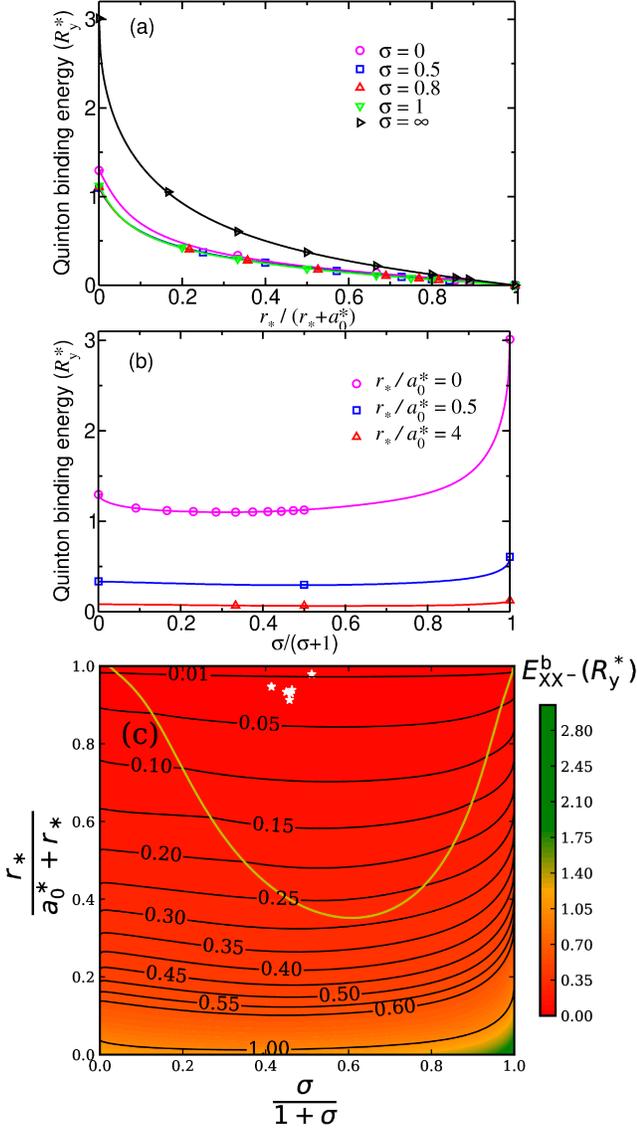}
\caption{(a) DMC BEs of XX$^-$ as a function of $r_*/(r_*+a_0^*)$. (b) DMC BEs of XX$^-$ as a function of $\sigma/(\sigma+1)$. (c) XX$^-$ BEs as a function of rescaled susceptibility and mass ratio. Above the yellow line $E_{{\rm XX}^-}^{\rm DE}<$ XX electron affinity, so that X and X$^-$ are the most energetically competitive.  Below the yellow line the situation is reversed, so that XX and free e are the most competitive. The white stars show the mass ratios and in-plane susceptibility of 1L-MoS$_2$ at (0.45,0.93), 1L-MoSe$_2$ at (0.46, 0.94), 1L-MoTe$_2$ at (0.50,0.98), 1L-WS$_2$ at (0.46,0.91), 1L-WSe$_2$ at (0.46,0.93), 1L-WTe$_2$ at (0.41,0.95), where the first and second numbers in brackets are $\sigma/(\sigma+1)$ and $r_*/(r_*+a_0^*)$, respectively. XX$^-$ BEs are between $0.00736(5)\,R_{\rm y}^*$ and $0.0288(1)\, R_{\rm y}^*$, with the numbers in brackets the BE error bars.\label{fig:be_quint}}
\end{figure}

Using the BE fits in Eqs.48,49 of Ref.\onlinecite{Mostaani_2017} together with Eqs.\ref{eq:af},\ref{eq:quint_be},\ref{eq:quint_deex}, we calculate the XX$^-$ BEs in Fig.\ref{fig:be_quint}. Above the yellow line in Fig.\ref{fig:be_quint}c, the XX$^-$ BE =$E_{{\rm XX}^-}^{\rm DE}$ (all 1L-TMDs fall in this region). Below the yellow line, the XX$^-$ BE is equal to the XX electron affinity.

Table \ref{table:compare_quint} lists the XX$^-$ BEs from DMC and the fit of Eq.\ref{eq:quint_deex} for 1L-WS$_2$, 1L-WSe$_2$, 1L-WTe$_2$, 1L-MoS$_2$, 1L-MoSe$_2$, 1L-MoTe$_2$.

To measure X$^-$, XX, XX$^-$ BEs, in Ref.\onlinecite{Barbone_2018} we used continuous wave (CW) PL at 4K for 1L-WSe$_2$ encapsulated between 2 10nm (bottom) and 3nm (top) ML-hBN on Si/SiO$_2$. Refs.\onlinecite{Paur_2019,LiT_2018,Barbone_2018} used different experimental conditions for 1L-WSe$_2$, but they all produced similar results, Table \ref{table:compare_quint}, across various techniques and substrates.

The size of a charge-carrier complex in a 1L-TMD can be defined by $r_0$. This is$\sim8${\AA} in TMDs listed in Table \ref{table:compare_quint}, because their e and h masses and screening lengths  are around the same order. Hence, we suggest that encapsulation in$>1$nm ML-hBN can be described by the permittivity $\epsilon=4\epsilon_0$\cite{Geick1966,Plass1997,Barth1998,Rumyantsev2001} of bulk hBN.

For ML-hBN-encapsulated TMDs we test 2 approaches to compare our results with experiments.

1) We fix $\epsilon=\epsilon_0$ and determine $r_*$ by fitting Eqs.48,49 of Ref.\onlinecite{Mostaani_2017} to the experimental X$^-$, XX BEs in Refs.\onlinecite{Hao_2017,Paur_2019,LiT_2018,Barbone_2018}. This is reasonable because, at distances larger than the layer-layer separation, the Keldysh interaction of Eq.\ref{eq:v_pot} for a ML is of the same form as for a 1L\cite{Danovich2018}, but with $r_*$ being the sum of $r_*$ for the different layers\cite{Danovich2018}. For $r\ll r_*$, only $\epsilon r_*$ appears in the logarithmic approximation to the Keldysh interaction in Eq.\ref{eq:log_int}, apart from a constant contribution to the total energy, which cancels out of $E_{{\rm XX}^-}^{\rm DE}$. Hence, it is preferable to fix $\epsilon$, and treat $r_*$ as the independent parameter. The XX$^-$ BEs calculated with this approach for 1L-WSe$_2$ and 1L-WS$_2$ encapsulated in ML-hBN agree with the experiments in Refs.\onlinecite{Barbone_2018, Paur_2019}, differing by at most$\sim2$meV, as for Table \ref{table:compare_quint}.

2) We use \textit{ab initio} vacuum $r_*$ for 1L-TMDs. To describe hBN encapsulation we use $\epsilon=4\epsilon_0$, consistent with Refs.\onlinecite{Geick1966,Plass1997,Barth1998,Rumyantsev2001}. This gives BEs$\sim$5-18meV smaller than Refs.\onlinecite{Barbone_2018,Paur_2019}. This difference could either due to the phenomenological parameters ($r_*$ and $\epsilon$), obtained by \textit{ab initio} methods and used in the Mott-Wannier-Keldysh model of Eq.\ref{eq:sch}, or be a result of neglecting intervalley scattering\cite{Dery2016} and contact (exchange) interactions.

The substrate effect on the BE of a charge-carrier complex can be described by:
\begin{equation}
\epsilon=(\epsilon_0+\epsilon_{\rm substrate})/2,
\label{eq:eff_eps}
\end{equation}
where $\epsilon_{\rm substrate}$ is the bulk permittivity of the substrate. In Ref.\onlinecite{LiT_2018}, 1L-WSe$_2$ was grown on SiO$_2$ by physical vapor deposition (PVD) and the X$^-$ and XX$^-$ BEs of 1L-WSe$_2$ at RT were measured as$\sim$30 and 51meV, respectively, by CW PL. Using the permittivity of SiO$_2$, $\epsilon_{\rm substrate}\sim 3.9\epsilon_0$\cite{Murase1994} in Eqs.\ref{eq:eff_eps}, \ref{eq:quint_deex}, and Eq.49 of Ref.\onlinecite{Mostaani_2017}, we calculate the X$^-$ and XX$^-$ BEs to be$\sim$19 and 38meV, respectively,$\sim$10-13meV less than the experiments in Ref.\onlinecite{LiT_2018} done at RT, while our calculations correspond to T=0K. Approximating the sapphire permittivity as isotropic with $\epsilon=8.9\epsilon_0$\cite{Harman_1994} gives X$^-$, XX, XX$^-$ BEs in 1L-MoSe$_2$ to be$\sim$13.6, 13.7, 30meV, respectively. In Ref.\onlinecite{Hao_2017} exfoliated 1L-MoSe$_2$ was transferred to a sapphire substrate and 2dCS at 13K was used to measure X$^-$, XX, and XX$^-$ BEs$\sim$27,18, 40meV, respectively\cite{Hao_2017}. Substrate-induced roughness can also cause inhomogeneity in the electronic structure and extra carrier scattering\cite{Chae2017}. This affects PL, leading to inhomogeneous broadening\cite{Ajayi_2017,Shree_2019}, which makes it difficult to identify charge-carrier complexes\cite{Hao_2017}. In Ref.\onlinecite{Hao_2017}, PL spectra were not recorded as a function of excitation power. However, power-dependent measurements help assign the PL peaks to X$^-$, XX, XX$^-$, because they show, respectively, sublinear, quadratic, and superlinear dependence with excitation power.

Due to the complexity in defining $r_*$ and $\epsilon$ for 1L-TMDs encapsulated in hBN or placed on a substrate, we use the first approach, where we fix $\epsilon=\epsilon_0$ and determine $r_*$ by fitting theoretical X$^-$ and XX BEs to available experiments, so to define the XX$^-$ BEs.
\subsection{Other charge-carrier complexes \label{sec:galaxies}}
We investigate doubly charged complexes, including XX$^{2-}$ (4e and 2h) and D$^0$hh (one positive donor ion, 1e, 2h), in which all charge carriers are distinguishable. Optimizing wave functions with pairwise and three-body correlations by variational Monte Carlo (VMC) energy minimization\cite{Umrigar_2007} does not result in bound-state wave functions. If we constrain the wave function to be bound, and then perform DMC, the resulting energy confirms that complexes are unbound. Thus, doubly charged complexes are unstable for all relevant material parameters.

In Ref.\onlinecite{Mostaani_2017} we considered what is the largest stable charge-carrier complex that can occur in 1L-TMDs. We showed that XX with two indistinguishable e is unstable in 1L-TMDs, because of the resulting antisymmetry of the spatial wave function. We concluded that, in bound complexes featuring only singly charged dopant ions and charge carriers, all charge carriers must be distinguishable. Our results show that a charge-carrier complex can feature at most one dopant ion. Because of the band structure (see Fig.\ref{fig:quint_class}a), 1L-TMDs can have 4e species and 2h species. This suggests that the largest stable cluster will have a positive donor ion, 4 distinguishable e, and 2 distinguishable h. We get bound-state wave functions describing the donor-bound double-negative XX (${\rm D}^{-}{\rm XX}$). These seven-body complexes are predicted to be stable in 1L-WS$_2$, 1L-WSe$_2$, 1L-MoS$_2$, 1L-MoSe$_2$ in vacuum and air. The DMC-calculated BEs with respect to the most energetically favorable products [donor-bound negative X (${\rm D}^{0}{\rm X}^{-}$)+free X] are in Table \ref{table:dbdnb}. Because the dominant decay products include an X, the BE gives the PL peak position of the ${\rm D}^{-}{\rm XX}$ complex relative to the X line, possible in samples containing donor defects.
\begin{table}
\centering
\caption{Theoretical $E^{\rm b}_{\rm D^-XX}$ for 1L-TMDs in vacuum, with \textit{ab initio} masses and $r_*$ from Table \ref{table:compare_quint}}
\begin{tabular}{lc}
\hline\hline
TMD & $E^{\rm b}_{\rm D^-XX}$ (meV) \\
\hline
1L-MoS$_2$ (vac.) & $58.3(5)$ \\
1L-MoSe$_2$ (vac.) & $78.6(3)$ \\
1L-WS$_2$ (vac.) & $80.4(5)$ \\
1L-WSe$_2$ (vac.) & $70.5(7)$ \\
\hline\hline
\end{tabular}
\label{table:dbdnb}
\end{table}
\subsection{Accuracy of the Rytova-Keldysh interaction}
\begin{figure}
\centerline{\includegraphics[width=90mm]{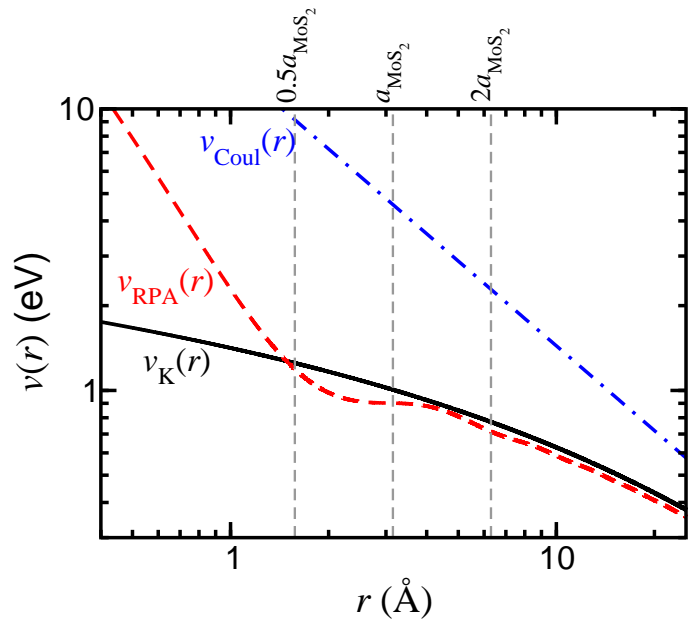}}
\caption{Unscreened Coulomb interaction potential: $v_{\rm Coul}(r)=1/r$; RKI: $v_{\rm K}(r)=V(r/r_*)/r_*$, with $r_*$ of 1L-MoS$_2$ in vacuum from Table \ref{table:compare_quint}; RPAI: $v_{\rm RPA}(r)$. $a_{\text{1L-MoS}_2}=3.15${\AA}\cite{Kormanyos_2015}}
\label{fig:aml_int}
\end{figure}

RKI arises from the approximation that the in-plane susceptibility of a material is a constant\cite{Rytova_1965}. Here, we investigate the RKI accuracy by using an alternative approach based on \textit{ab initio} calculations for 1L-MoS$_2$. Realistic dielectric functions exhibit spatial dependencies which differ from the Coulomb interaction at short range ($r\ll r_*$). At long range ($r\gg r_*$), in-plane screening becomes irrelevant, and all physical dielectric functions behave as the Coulomb interaction, as explained in Methods. Given that the binding of excitonic complexes occurs on length scales larger than the lattice spacings (Table \ref{table:compare_quint}), where screening effects are most prominent (Fig.\ref{fig:aml_int}), an investigation into their effects on charge-carrier binding is warranted. Ref.\onlinecite{Aghajanian_2018} parameterized a dielectric permittivity $\epsilon({\bf q})$ for 1L-MoS$_2$ via RPA applied to Kohn-Sham orbitals from density functional theory calculations to study charged defects. We refer to the real-space interaction formed from $\epsilon({\bf q})$ as the RPA interaction (RPAI), and compare it to RKI in Fig.\ref{fig:aml_int}.
\begin{figure}
\centerline{\includegraphics[width=90mm]{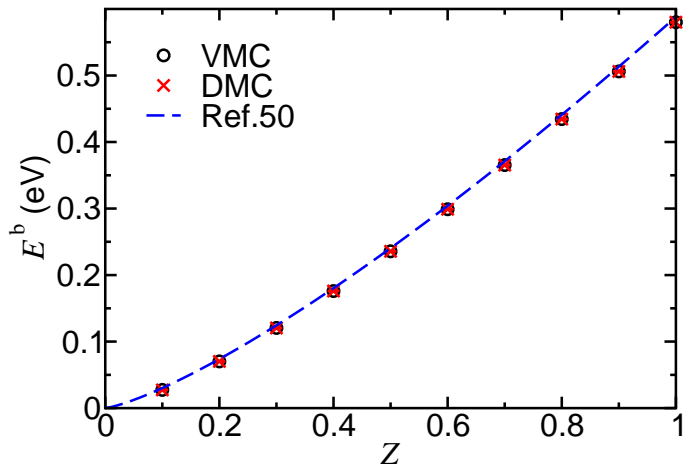}}
\caption{BE of a donor atom in 1L-MoS$_2$ as a function of donor impurity charge $Z$. VMC and DMC results are compared with the numerical data of Ref.\onlinecite{Aghajanian_2018}. \label{fig:aml_comp}}
\end{figure}

We use cusp conditions\cite{Needs_2020} to prevent the wave function of charge carriers to diverge around particle coalescence points\cite{Mostaani_2017}. We use the same trial-wave-function form as our calculations with RKI, see Methods for details. As a test, Fig.\ref{fig:aml_comp} verifies we reproduce the theoretical donor-atom BEs of Ref.\onlinecite{Aghajanian_2018}, for the case of an adatom-bound e above a 1L-MoS$_2$ surface. Our data have small$\sim10^{-4}-10^{-3}$meV error bars, and differ from Ref.\onlinecite{Aghajanian_2018} by a few meV, for typical BEs$\sim$few hundreds meV.

Fig.\ref{fig:aml_int} indicates that at distances$>a_{\text{1L-MoS}_2}$, RPAI follows the same form as RKI (and, ultimately, Coulomb) interactions. However, at distances$\sim a_{\text{1L-MoS}_2}$, RKI no longer overlaps RPAI, hence cannot describe the interaction between quasiparticles. Fig.\ref{fig:aml_int} shows that for $r\ll 0.5 a_{\text{1L-MoS}_2}$, RPAI reduces to an unscreened $1/r$ potential, while the RKI behavior is that of a logarithmic divergence\cite{Keldysh_1979}. However, within the effective-mass approximation, we can only describe quasiparticles on length scales$>a_{\text{1L-MoS}_2}$, as shown in Table \ref{table:compare_quint} and Fig.\ref{fig:aml_int}, whose associated Bloch wave packets are localized in momentum space, with well-defined effective mass.

The RPAI BEs of charge-carrier complexes are in Table \ref{table:RPA_data}. Removing the bare Coulomb interaction at distances$<a_{\text{1L-MoS}_2}$ is necessary to obtain results in agreement with previous experimental\cite{ Feng_2012,Li_2015} and theoretical\cite{Lin_2014} works. For $r_{\rm c}<a_{\text{1L-MoS}_2}$, we truncate the RPAI to a constant $v(r<r_{\rm c})=v(r_{\rm c})$. The precise value of $r_{\rm c}$ is not particularly important for the BE calculation of charge-carrier complexes, as we observe a weak BE dependence on this parameter, see Table \ref{table:RPA_data}.
\begin{table}
\centering
\caption{BEs of charge-carrier complexes in 1L-MoS$_2$ calculated using different interaction potentials. R-RPA is the rounded RPAI, with $r_{\rm c}$ in brackets. RKI values from Ref.\onlinecite{Mostaani_2017}.}
\begin{tabular}{lr@{}lr@{}lr@{}l}
\hline\hline
\multirow{2}{*}{Interaction potential} & \multicolumn{6}{c}{BE (meV)}
\\
& \multicolumn{2}{c}{{\rm X}} & \multicolumn{2}{c}{{\rm X}$^{-}$} &
\multicolumn{2}{c}{{\rm XX}} \\
\hline
Bare RPA & 765&.45(2) & 77&.7(7) & 184&.5(7) \\
R-RPA ($2\times a_{\text{MoS}_2}$)& 454&.5(1) & 30&.0(3) & 16&.0(4)\\
R-RPA ($a_{\text{MoS}_2}$) & 483&.8(4) & 30&.7(4) & 19&.4(8) \\
R-RPA ($0.5\times a_{\text{MoS}_2}$)& 492&.47(3) & 30&.9(2) &
20&.4(4)\\
RKI & 546&.5 & 35&.0 & 23&.5 \\
\hline
Experiment       & ~~$\sim 500$& ~\cite{Feng_2012,Li_2015} & & & \\
$GW$-BSE         &             &                           & 40&.0
\cite{Lin_2014}& & \\
\hline\hline
\end{tabular}
\label{table:RPA_data}
\end{table}

Table \ref{table:RPA_data} indicates that there is no need to use an expression for the electrostatic interaction between charge carriers in LSMs more sophisticated than RKI when evaluating BEs of trions, biexcitons, and quintons. As explained in Methods, any errors in the Mott-Wannier-Keldysh model of charge-carrier complexes for an isolated 1L are either due to the parameters (effective masses, $r_*$, environment permittivity), or to a more fundamental breakdown of the effective-mass approximation. Intervalley scattering may play an important role in the complexes' BEs\cite{Aghajanian_2018}, while exchange effects could be relevant in highly localized complexes\cite{Elliot1961}.
\subsection{Complexes in uniform magnetic fields}
\label{sec:excitons}
For an out-of-plane external magnetic field of flux density ${\bf B}=(0,0,B)$, where $B$ is a positive constant, we can write the Hamiltonian as:
\begin{align}
\hat{H} & = \sum_i \frac{1}{2m_i}{(-i\hbar\nabla_i-q_i{\bf A}_i)}^2+\sum_{i>j}\frac{q_i q_j}{4\pi\epsilon r_*}V(r_{ij}/r_*)\nonumber \\ & = \sum_i \left(-\frac{\hbar^2}{2m_i}\nabla_i^2+i\frac{\hbar q_i}{m_i} {\bf A}_i \cdot \nabla_i +\frac{q_i^2 |{\bf A}_i|^2}{2m_i} \right) \nonumber \\ &
\qquad {} +\sum_{i>j}\frac{q_i q_j}{4\pi\epsilon r_*}V(r_{ij}/r_*)
\label{eq:hamiltonian}
\end{align}
where ${\bf A}_i=-{\bf r}_i \times {\bf B}/2=(-y_i,x_i,0)B/2$ is the magnetic vector potential for particle $i$ in the Coulomb gauge (so that $\nabla_i \cdot {\bf A}_i=0$)\cite{Griffiths2016}. We neglect the charge carriers' intrinsic magnetic dipole moment energy in the external magnetic field, because this contribution cancels out.

Substituting ${\bf A}_i$ into Eq.\ref{eq:hamiltonian}, the term $q_i^2|{\bf A}_i|^2/(2m_i)=q_i^2B^2|{\bf r}_i|^2/(8m_i)$ provides a quadratic confining potential for the particles in the complex. This cannot be regarded as a perturbation for the (otherwise free) center-of-mass (CoM) motion, because there is a quantitative difference between a bound state wave function in a quadratic potential and free motion in zero potential, no matter how small the quadratic coefficient\cite{Kittel2004}. The zero-point energy of the CoM motion in the confining potential results in a linear [$O(B)$] contribution to the total energy, as given in Eq.\ref{eq:com_zpe}. The term also weakly perturbs the relative motion within the complex, giving a quadratic [$O(B^2)$] contribution to the energy. We thus include the $q_i^2|{\bf A}_i|^2/(2m_i)=q_i^2B^2|{\bf r}_i|^2/(8m_i)$ term in our QMC calculations. The linear $(i\hbar q_i/m_i) {\bf A}_i \cdot \nabla_i$ term in Eq.\ref{eq:hamiltonian} breaks time-reversal symmetry as it is imaginary\cite{Sachs1987}. It only adds to the energy in second-order perturbation theory, giving another $O(B^2)$ contribution. This vanishes when we use a variational Ansatz consisting of a real trial wave function. We therefore neglect it.

The ground-state energies of isolated e/h are $E_{\rm e}=\hbar eB/(2 m_{\rm e})$ and $E_{\rm h}=\hbar eB/(2 m_{\rm h})$, in the presence of a magnetic field\cite{Griffiths2016}. More generally, if a bound complex of $N_{\rm e}$ e and $N_{\rm h}$ h moves in a magnetic field, from Eq.\ref{eq:hamiltonian} the quadratic confining potential is:
\begin{equation}
U=\sum_i \frac{e^2B^2|{\bf r}_i|^2}{8m_i}\approx\frac{B^2e^2}{8}\left(\frac{N_{\rm e}}{m_{\rm e}}+\frac{N_{\rm h}}{m_{\rm h}}\right) R^2,
\end{equation}
where ${\bf R}$ is the CoM position. The total mass of the complex is $N_{\rm e}m_{\rm e}+N_{\rm h}m_{\rm h}$. Hence, we obtain the CoM zero-point energy of a charge complex as:
\begin{equation}
E^{\rm CoM}=\frac{\hbar eB}{2}\sqrt{\frac{N_{\rm e}/m_{\rm e}+N_{\rm h}/m_{\rm h}}{N_{\rm e}m_{\rm e}+N_{\rm h}m_{\rm h}}}.
\label{eq:com_zpe}
\end{equation}
If $m_{\rm e}=m_{\rm h}\equiv m$ then $E^{\rm CoM}=\hbar eB/(2m)$, independent of $N_{\rm e}$, $N_{\rm h}$. For a bound complex, our results show that the magnetic field can always be made sufficiently weak so that the external potential is slowly varying on the length scale of the complex (i.e. $\sqrt{\hbar/(eB)}<a_0^\ast$). Hence, Eq.\ref{eq:com_zpe} is the leading-order contribution to the free charge-carrier complex energy in a magnetic field.
\begin{figure*}
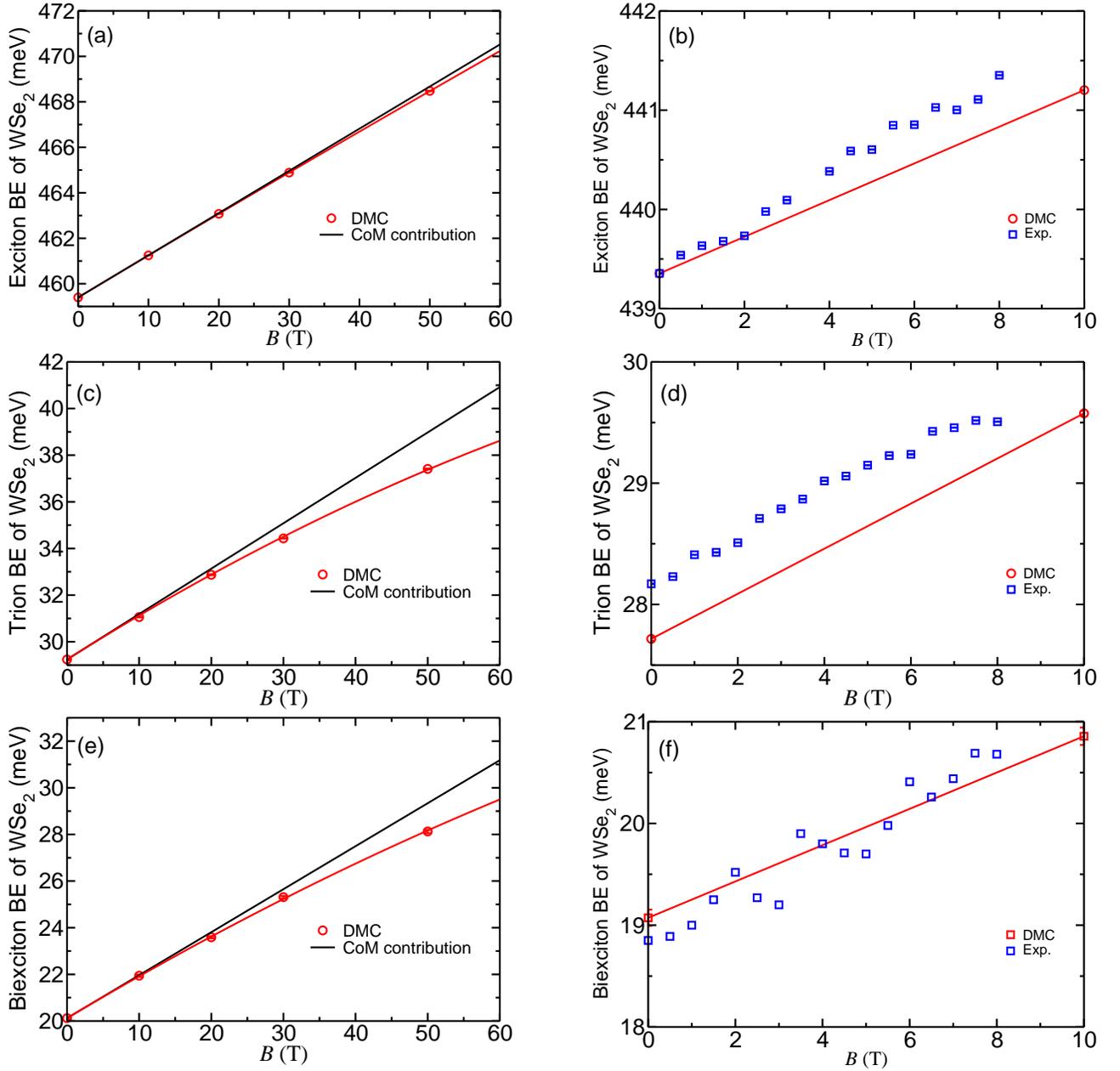

  \begin{minipage}{0.49\textwidth}
    \centering \includegraphics[width=.9\linewidth]{ex_bewse2.eps}
  \end{minipage}\hfill
  \begin{minipage}{0.49\textwidth}
    \centering
    \includegraphics[width=.9\linewidth]{ex_bewse2_exptheo.eps}
  \end{minipage}
  \begin{minipage}{0.49\textwidth}
    \centering \includegraphics[width=.9\linewidth]{trion_bewse2.eps}
  \end{minipage}\hfill
  \begin{minipage}{0.49\textwidth}
    \centering
    \includegraphics[width=.9\linewidth]{trion_bewse2_exptheo.eps}
  \end{minipage}
  \begin{minipage}{0.49\textwidth}
    \centering
    \includegraphics[width=.9\linewidth]{biexciton_bewse2.eps}
  \end{minipage}\hfill
  \begin{minipage}{0.49\textwidth}
    \centering
    \includegraphics[width=.9\linewidth]{biexciton_bewse2_exptheo.eps}
  \end{minipage}
\caption{Theoretical BEs of (a) X, (c) X$^-$, and (e) XX as a function of perpendicular magnetic field for 1L-WSe$_2$ in vacuum. We use the \textit{ab initio} mass and $r_*$ parameters of Table \ref{table:compare_quint}. The CoM contribution for X is $E_{\rm X}^{\rm b,CoM}={(E_{\rm X}^{\rm b})}_{B=0}+E_{\rm e}+E_{\rm h}-E_{\rm X}^{\rm CoM}$; for X$^-$ is $E_{\rm X^-}^{\rm b,CoM}={\left(E_{\rm X^-}^{\rm b}\right)}_{B=0}+E_{\rm e}-E_{\rm X^-}^{\rm CoM}$; and for XX is $E_{\rm XX}^{\rm b,CoM}={\left(E_{\rm XX}^{\rm b}\right)}_{B=0}+2E_{\rm X}^{\rm CoM}-E_{\rm XX}^{\rm CoM}$. Experimental BEs of (b) X, (d) X$^-$, (f) XX for 1L-WSe$_2$ encapsulated in hBN, compared with DMC ones using $\epsilon=\epsilon_0$ and $r_*=48${\AA} and the fit to Eq.\ref{eq:CBfit}\label{fig:wse2}}
\end{figure*}

Fig.\ref{fig:wse2} plots the DMC X, X$^-$, XX BEs for 1L-WSe$_2$ in vacuum, in the presence of an out-of-plane magnetic field, using RKI. $m_{\rm e}$, $m_{\rm h}$, and $r_*$ are taken from Table \ref{table:compare_quint}. Our results are in agreement with Ref.\onlinecite{Donck_2018}. The CoM contribution of Eq.\ref{eq:com_zpe} is a good approximation to calculate the X, X$^-$, XX BEs in magnetic fields$<8$T, because it is exact up to linear order in magnetic field within the effective-mass approximation. For the X BE in magnetic fields$>8$T, we use Eq.\ref{eq:CBfit}, derived in Methods. The fitted $C$ in Eq.\ref{eq:CBfit} is 0.557 for X in 1L-WSe$_2$.

Figs\ref{fig:wse2}b,d,f compare our DMC BEs with measurements for hBN-encapsulated 1L-WSe$_2$. The sample is produced by exfoliating flux zone grown B-WSe$_{2}$\cite{Zhangx_2015}, then encapsulating it with ML-hBN (10nm bottom and 3nm top) using an all-dry technique\cite{Bonaccorso2012,Purdie2018}. Measurements are done in a closed-cycle cryostat (Attocube Attodry 1000) at 4K with superconducting magnets allowing out-of-plane magnetic fields up to 8T. CW excitation is provided with a diode laser at 658nm, close to the 1L-WSe$_2$ optical band gap\cite{Keliang2014}. Polarization-resolved excitation and collection pass through a confocal microscope with the sample in reflection geometry. The PL signal is sent to a liquid-N$_2$-cooled spectrometer (Princeton).

We assume $r_*=48$ {\AA} and $\epsilon=\epsilon_0$, as discussed in Sec.\ref{sub_be_quint}. The theoretical and experimental BEs differ$<0.3$meV over the 0-8K temperature range. The $O(B)$ magnetic-field dependence is only via the effective masses and $N_\text{e}$, $N_\text{h}$, via the CoM energy, Eq.\ref{eq:com_zpe}. The fact that the theoretical and experimental magnetic-field trends in Fig.\ref{fig:wse2} agree well demonstrates that the approximation with \textit{ab initio} effective masses is accurate. The main challenge is to obtain a sufficiently accurate interaction between charge carriers. The BE $O(B)$ term is the same for all complexes, in the limit $m_{\rm e}=m_{\rm h}$. For most 1L-TMDs, $m_{\rm e}$ and $m_{\rm h}$ are similar, Table \ref{table:compare_quint}, implying that the magnetic-field dependence cannot be used to distinguish carrier complexes. Table \ref{table:bewse2_magnetic} has DMC and experimental X, X$^-$, XX BEs for 1L-WSe$_2$ in the presence of an out-of-plane external magnetic field, as for Fig.\ref{fig:wse2}. The variation of BEs of different charge complexes is the same$<8$T.
\begin{table}
\centering
\caption{DMC and experimental X, X$^-$, XX BEs in meV for 1L-WSe$_2$ with an out-of-plane external magnetic field.}
\begin{tabular}{lr@{}lr@{}lr@{}lr@{}lr@{}lr@{}l}
\hline\hline
\multirow{2}{*}{B (T)} &\multicolumn{6}{c}{DMC} &
\multicolumn{6}{c}{Experiment}\\
& \multicolumn{2}{c}{X} & \multicolumn{2}{c}{X$^-$} &
\multicolumn{2}{c}{XX} & \multicolumn{2}{c}{X} &
\multicolumn{2}{c}{X$^-$} & \multicolumn{2}{c}{XX}\\
\hline
0 &$439.$&$35$ & $27.$&$72$& $19.$&$07$ & $439.$&$35$ & $28.$&$17$ & $18.$&$85$  \\
1 &$439.$&$54$ & $27.$&$90$& $19.$&$25$ & $439.$&$63$ & $28.$&$41$ & $19$&$ $    \\
2 &$439.$&$72$ & $28.$&$09$& $19.$&$43$ & $439.$&$73$ & $28.$&$51$ & $19.$&$52$  \\
3 &$439.$&$91$ & $28.$&$27$& $19.$&$61$ & $440.$&$09$ & $28.$&$79$ & $19.$&$2$   \\
4 &$440.$&$09$ & $28.$&$46$& $19.$&$79$ & $440.$&$38$ & $29.$&$02$ & $19.$&$8$   \\
5 &$440.$&$28$ & $28.$&$65$& $19.$&$96$ & $440.$&$60$ & $29.$&$23$ & $19.$&$7$   \\
6 &$440.$&$46$ & $28.$&$83$& $20.$&$14$ & $440.$&$85$ & $29.$&$24$ & $20.$&$41$  \\
7 &$440.$&$65$ & $29.$&$02$& $20.$&$32$ & $441.$&$00$ & $29.$&$46$ & $20.$&$44$  \\
8 &$440.$&$83$ & $29.$&$20$& $20.$&$5$  & $441.$&$35$ & $29.$&$51$ & $20.$&$68$  \\
\hline\hline
\end{tabular}
\label{table:bewse2_magnetic}
\end{table}
\begin{figure}[!htbp]
\centerline{\includegraphics[width=90mm]{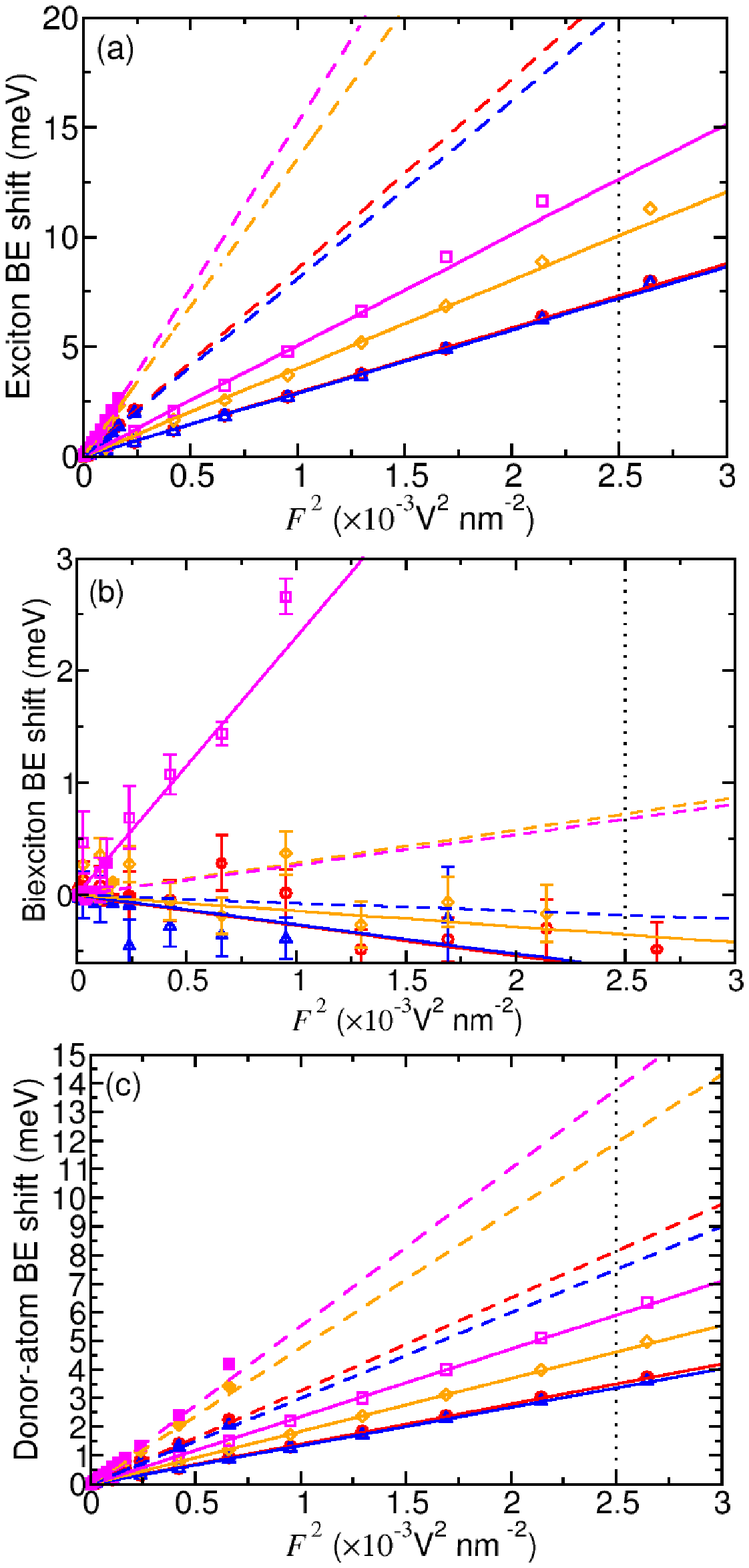}}
\centerline{\includegraphics[width=60mm]{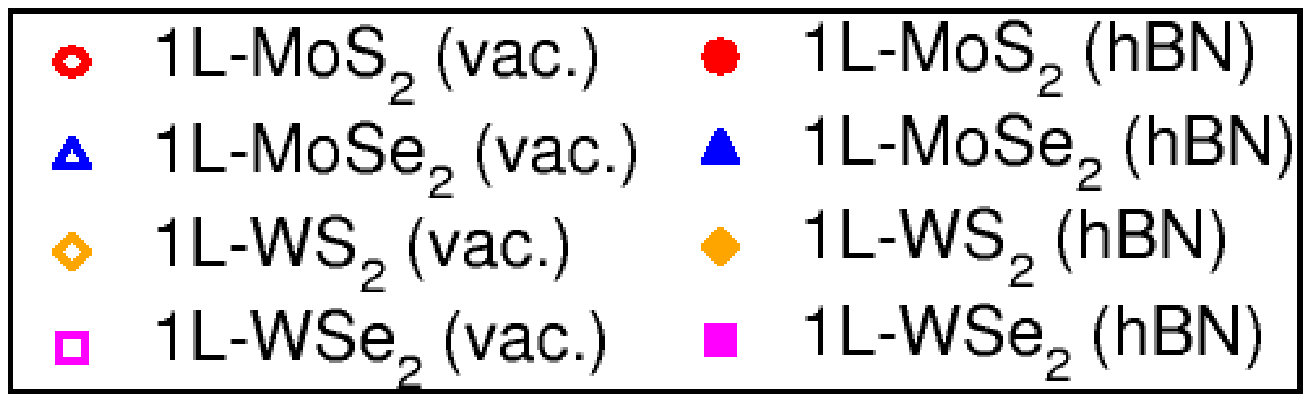}}
\caption{DMC BE shift for (a) X, (b) XX, (c) donor atoms as a function of $F^2$ for different 1L-TMDs in vacuum and encapsulated in hBN. Error bars in (a,c) are smaller than the symbols. The solid and dashed lines are BEs determined by the polarizabilities in Table \ref{table:polarizabilities} for 1L-TMDs in vacuum and encapsulated by hBN. The vertical dotted lines correspond to $F=50$mV\,nm$^{-1}$, beyond which VMC energy minimization does not result in bound-state wave functions.\label{fig:electricfield}}
\end{figure}
\subsection{Complexes in uniform electric fields}
A bias voltage $\Delta V$ applied to a 1L-LSM results in an in-plane electric field. Its precise form depends on device geometry. Here, we assume a uniform electric field $F=-\Delta V/d$, where $d$ is the distance between terminals, for simplicity. $F$ will perturb the energies of charge-carrier complexes in the CoM frame. We therefore investigate the effects of $F$ on BEs by including an additional term $-\sum_i q_i F x_i$ in the Hamiltonian, where $x_i$ is the $x$ coordinate of particle $i$. Fig.\ref{fig:electricfield} plots the X BE shift as a function of electric field strengths for 1L-MoS$_2$, 1L-MoSe$_2$, 1L-WS$_2$, 1L-WSe$_2$, in vacuum and encapsulated by hBN, using the \textit{ab initio} parameters in Table \ref{table:compare_quint} and $\epsilon=4\epsilon_0$. In each case, X BE goes as the square of the in-plane electric field, as expected for a linearly polarizable exciton\cite{Reitz2009}. Thus, the total energy of an isolated neutral complex of polarizability $\alpha$ in a uniform $F$ is:
\begin{equation} E=E_{F=0} - \alpha F^2/2,
\label{eq:electricfieldshift}
\end{equation}
where $E_{F=0}$ is the energy of the complex in the absence of external fields. The variation of energy with electric field strength remains quadratic up to at least$\sim50$mV\,nm$^{-1}$. $>50$ mV\,nm$^{-1}$ we find that optimizing wave functions by VMC energy minimization does not result in bound-state wave functions. If the parameters in the wave function are fixed such that a bound state is forced, the resulting DMC calculations are unstable. It is possible that some, or all, complexes remain bound at these larger electric fields, and our QMC calculations become unstable, due to the choice of trial wave function. The form we use is isotropic, so it does not allow the complex to polarize in VMC. Polarization arises at DMC level.
\begin{figure}[!htbp]
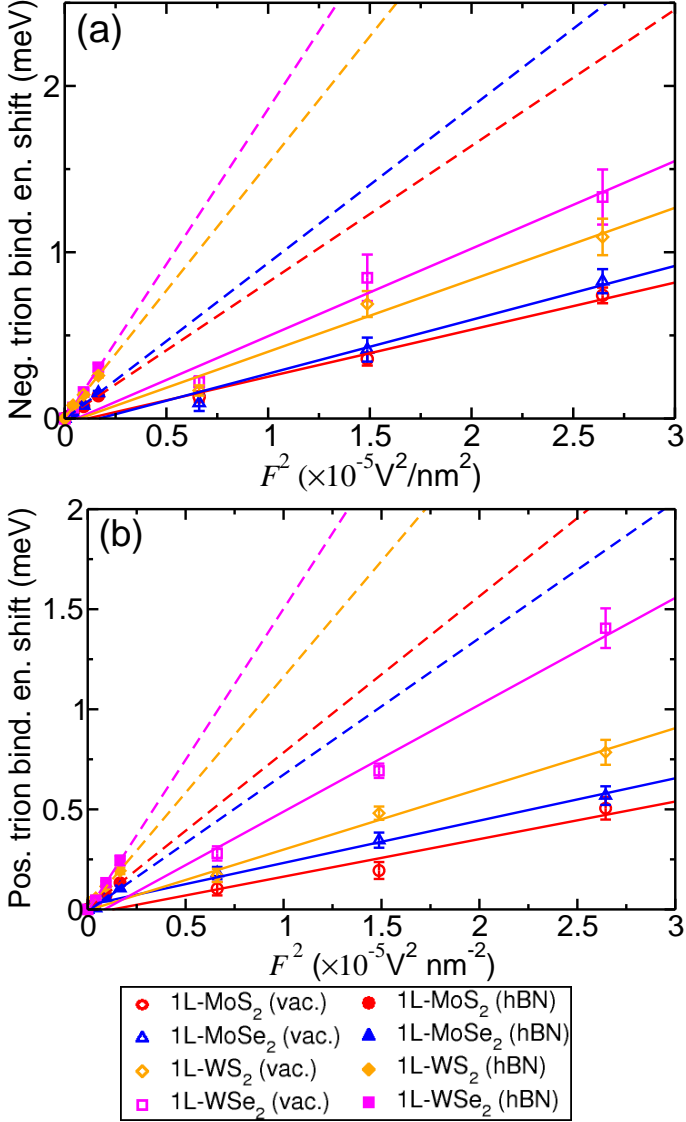

\centerline{\includegraphics[width=90mm]{Xminus_Efield_shift.eps}}
\centerline{\includegraphics[width=90mm]{Xplus_Efield_shift.eps}}
\centerline{\includegraphics[width=60mm]{Efield_legend.eps}}
\caption{DMC BE shift for (a) X$^-$ and (b) X$^+$ as a function of $F^2$ for different 1L-TMDs in vacuum and encapsulated in hBN. Where error bars are not visible they are smaller than the symbols. The solid and dashed lines show BEs determined by the polarizabilities in Table \ref{table:polarizabilities}\label{fig:trionelectricfield}}
\end{figure}
\begin{table}
\centering
\caption{Theoretical in-plane polarizabilities of X, XX, D$^0$, X$^-$, X$^+$ in 1L-TMDs, in vacuum and hBN encapsulated}
\begin{tabular}{lr@{}lr@{}lr@{}lr@{}lr@{}l}
\hline\hline
\multirow{2}{*}{TMD} & \multicolumn{10}{c}{Polarizability (eV\,nm$^2$\,V$^{-2}$)} \\
& \multicolumn{2}{c}{X} & \multicolumn{2}{c}{XX} & \multicolumn{2}{c}{D$^0$} & \multicolumn{2}{c}{X$^-$} & \multicolumn{2}{c}{X$^+$} \\
\hline
1L-MoS$_2$ (vac.)  & $5$&$.84(2)$ & $11$&$.14(8)$ & $2$&$.802(9)$ & $66$&$(6)$ & $44$&$(6)$ \\
1L-MoSe$_2$ (vac.) & $5$&$.76(2)$ & $11$&$.0(1)$ & $2$&$.687(9)$ &$80$&$(9)$ & $45$&$(6)$ \\
1L-WS$_2$ (vac.)   & $8$&$.04(3)$ & $15$&$.8(1)$ & $3$&$.70(1)$ &$108$&$(10)$  & $72$&$(7)$ \\
1L-WSe$_2$ (vac.)  & $10$&$.10(4)$ & $24$&$.8(3)$ & $3$&$.96(1)$ &$130$&$(16)$ & $118$&$(9)$ \\
1L-MoS$_2$ (hBN)  & $17$&$.17(4)$ & $34$&$.2(3)$ & $6$&$.51(2)$ &$179$&$(17)$ & $161$&$(22)$ \\
1L-MoSe$_2$ (hBN) & $16$&$.22(4)$ & $32$&$.3(2)$ & $6$&$.89(2)$ &$211$&$(22)$ & $181$&$(23)$ \\
1L-WS$_2$ (hBN)   & $27$&$.16(4)$ & $54$&$.9(3)$ & $4$&$.95(1)$ &$316$&$(27)$ & $246$&$(32)$ \\
1L-WSe$_2$ (hBN)  & $30$&$.43(4)$ & $61$&$.4(3)$ & $5$&$.29(1)$ &$409$&$(32)$ & $367$&$(32)$ \\
\hline\hline
\end{tabular}
\label{table:polarizabilities}
\end{table}
\begin{table}[!htbp]
\centering
\caption{Calculated BE shifts of X, XX, D$^0$, X$^{-}$, X$^{+}$ using Eq.\ref{eq:electricfieldshift} and polarizabilities in Table \ref{table:polarizabilities} for 1L-TMDs, both in vacuum and encapsulated by hBN, for $F=50$mV\,nm$^{-1}$. Not all complexes are bound at $F=50$ mV\,nm$^{-1}$}
\begin{tabular}{lr@{}lr@{}lr@{}lr@{}lr@{}l}
\hline\hline
\multirow{2}{*}{TMD} & \multicolumn{10}{c}{Binding-energy shift (meV)}
\\
& \multicolumn{2}{c}{X} & \multicolumn{2}{c}{XX}
&\multicolumn{2}{c}{~~D$^0$} & \multicolumn{2}{c}{~~~X$^-$}
&\multicolumn{2}{c}{~~~X$^+$} \\
\hline
1L-MoS$_2$ (vac.) & $7$&$.3$ & ~~$<1$& & ~~$3$&$.5$ & ~~~~$76$& &~~~~$48$& \\
1L-MoSe$_2$ (vac.) & $7$&$.2$ & $<1$& & $3$&$.4$ & $93$& & $49$& \\
1L-WS$_2$ (vac.) & $10$&$.1$ & $<1$& & $4$&$.6$ & $125$& & $80$& \\
1L-WSe$_2$ (vac.) & $12$&$.6$ & $5$&$.8$ & $4$&$.9$ & $150$& & $135$&
\\
1L-MoS$_2$ (hBN) & $21$&$.5$ & $<1$& & $8$&$.1$ & $201$& & $180$& \\
1L-MoSe$_2$ (hBN) & $20$&$.3$ & $<1$& & $8$&$.6$ & $243$& & $206$& \\
1L-WS$_2$ (hBN) & $34$&$.0$ & $<1$& & $6$&$.2$ & $362$& & $273$& \\
1L-WSe$_2$ (hBN) & $38$&$.0$ & $<1$& & $6$&$.6$ & $473$& & $408$& \\
\hline\hline
\end{tabular}
\label{table:expectedshifts}
\end{table}

The XX and donor-atom BEs vary linearly with $F^2$, Fig.\ref{fig:electricfield}. However, while the donor-atom BEs increase with $F^2$, the XX BEs decrease. For a 4-particle complex, alignment of charges in the direction of the applied field places like charges closer together, and reduces BE with respect to dissociation into two-particle complexes. Trion BEs also vary linearly with $F^2$, Fig.\ref{fig:trionelectricfield}. However, QMC calculations become unstable at much lower $F$. This is reflected in the higher polarizabilities for trions than for neutral complexes, Table \ref{table:polarizabilities}.

The predicted BE shifts of each of the complexes are in Table \ref{table:expectedshifts} for 1L-TMDs, both in vacuum and encapsulated by hBN, subject to $F=50$mV\,nm$^{-1}$, beyond which VMC energy minimization does not result in bound-state wave functions. The shifts in the peaks of the trions are so large that, at the very least, they should be experimentally distinguished from the neutral complexes when an electric field is applied. Identification of a positive from a negative trion may be possible in some materials/environments, but not all. For neutral complexes, the differences of a few meV in BE shifts suggest they are unlikely to be experimentally identified by their peak shifts under an electric field.
\section{Conclusions \label{sec:conclusions}}
We used DMC to calculate XX$^-$ BEs in 1L-LSMs within the effective-mass approximation, using the RKI potential. A program available online \cite{mostaani2022} can be used to evaluate interpolated XX$^-$ BEs given e and h effective masses, in-plane susceptibility, and environment permittivity for a desired 1L-LSM. The BEs of charge-carrier complexes in 1L-LSMs in vacuum from RKI are in excellent agreement with those obtained using interaction potentials taken from \textit{ab initio} RPA, suggesting RKI is a reliable interaction potential to describe screened interaction between charge carriers in 1L-LSMs.

We also considered the effect of external out-of-plane magnetic fields and in-plane electric fields on BEs of charge-carrier complexes in 1L-LSMs. The resulting BE changes are linear in magnetic fields and quadratic in electric fields up to 10T and 50mV\,nm$^{-1}$.

We measured X, X$^-$, XX BEs for hBN-encapsulated 1L-WSe$_2$ up to 8T, where the BEs vary linearly with magnetic field, and found them to be in good agreement with the effective-mass approximation using \textit{ab initio} effective masses. These BE shifts could in principle be used to identify complexes in PL experiments, provided $m_{\rm e}^*$ and $m_{\rm h}^*$ are different. In practice, $m_{\rm e}^*$ and $m_{\rm h}^*$ in 1L-TMDs are too similar to distinguish complexes in external magnetic fields. In-plane electric fields should shift the BE peaks in proportion to the field strength and allow for identification of charged from neutral complexes.

We derived BEs of charge-carrier complexes in 1L-TMDs by solving the interacting quantum few-body problem for each complex, working within the effective-mass approximation, with a RKI potential between charge carriers. The BE magnetic-field dependence agrees with experiments on a sub-meV energy scale. Since this only involves $m_{\rm e}^*$ and $m_{\rm h}^*$, and not the parameters describing the screened interaction, the approximation with \textit{ab initio} effective masses is highly accurate.

Efforts to improve the quantitative accuracy of BE calculations should therefore focus on the description of substrate and environmental screening, and on the inclusion of contact interactions and intervalley scattering.
\begin{acknowledgments}
We thank M. Aghajanian, A. A. Mostofi, J. Lischner, V. I. Fal'ko, G. Wang for useful discussions. We acknowledge support from EPSRC Grants EP/P010180/1, EP/L01548X/1, EP/K01711X/1, EP/K017144/1, EP/N010345/1, EP/L016087/1, ERC grants Corr-NEQM, Hetero2D, GSYNCOR, Lancaster University's High-End Computing facility, EU Graphene and Quantum Flagship. The data underlying this manuscript available at https://doi.org/10.17863/CAM.87211. For open access, we applied a Creative Commons Attribution (CC BY) license to any Author Accepted Manuscript version arising from this submission.
\end{acknowledgments}
\section{Methods }\label{sec:method}
\subsection{Effective-mass approximation} \label{subsec_effmass}
All our calculations are performed within the effective-mass approximation. For charge-carrier complexes in 1L-LSMs in the absence of external fields, we solve the Mott-Wannier-Keldysh Schr\"{o}dinger equation\cite{Mostaani_2017}:
\begin{equation}
\left[-\sum_i\frac{\hbar^2}{2m_i}\nabla_i^2+\sum_{i>j}\frac{q_i q_j}{4\pi\epsilon r_*} V(r_{ij}/r_*)\right]\psi=E\psi,
\label{eq:sch}
\end{equation}
where $m_i$ and $q_i$ are the band effective mass and charge of particle $i$, $r_{ij}$ is the separation of particles $i$ and $j$, $E$ is the energy eigenvalue, $\epsilon$ is the absolute permittivity of the surrounding medium, and $r_* \equiv \kappa/(2\epsilon)$, where $\kappa$ is the in-plane susceptibility. In Eq.\ref{eq:sch} the electrostatic interaction potential $V$, known as RKI\cite{Rytova_1965,Rytova_1967,Keldysh_1979}, is given by\cite{Mostaani_2017}:
\begin{equation}
V(r/r_*)=\frac{\pi}{2}\left[H_0(r/r_*)- Y_0(r/r_*)\right],
\label{eq:v_pot}
\end{equation}
where $H_n(x)$ is a Struve function\cite{Arfken2013} and $Y_n(x)$ is a Bessel function of the second kind\cite{Arfken2013}. At long range ($r\gg r_*$) the potential in Eq.\ref{eq:v_pot} is a Coulomb interaction $V(r/r_*)\sim r_*/r$; at short range ($r\ll r_*$), logarithmic:
\begin{equation}
V(r/r_*) \approx -\ln\left(\frac{e^\gamma r}{2r_*}\right),
\label{eq:log_int}
\end{equation}
where $\gamma \sim0.57721$ is Euler's constant\cite{Arfken2013}.

We do not include contact interactions between charge carriers due to exchange and correlation effects that occur when they are localized on the same site\cite{Elliot1961}, since these partially cancel out of BEs for complexes larger than X.
\subsection{QMC calculations}
\label{app:qmc}
We use VMC\cite{Umrigar_2007} and DMC\cite{Ceperley_1980,Foulkes_2001} to calculate the total energies of complexes of charge carriers in 1L-LSMs. We use the RKI potential in Eq.\ref{eq:v_pot} or, for the short range ($r\ll r_*$) limit, the logarithmic interaction of Eq.\ref{eq:log_int}. Our trial wave functions for complexes of distinguishable charge carriers are of the Jastrow form\cite{Needs_2020}, which includes a pairwise sum of terms depending on the distances between charge carriers, as for Ref.\onlinecite{Mostaani_2017}. Trial wave functions are optimized within VMC by minimizing first the energy variance\cite{Umrigar_1988,Drummond_2005}, then the energy expectation\cite{Umrigar_2007}. Our fixed-node DMC energies are exact solutions to the Mott-Wannier-Keldysh model of Eq.\ref{eq:sch}. DMC calculations use time steps in the ratio $1:4$, with the corresponding target configuration populations in the ratio $4:1$. The resulting energies are extrapolated linearly to zero time step and to infinite population. QMC calculations are done in the \textsc{casino} code\cite{Needs_2020}.
\subsection{Fitting function for BE as a function of magnetic field
\label{app:BE_in_Bfield}}
We consider a complex of $N_{\rm e}$ and $N_{\rm h}$ e and h interacting via the logarithmic approximation to the Keldysh interaction in the presence of a uniform magnetic field ${\bf B}=(0,0,B)$. Let $\tilde{B}=B/B_0$, $\tilde{m}_i=m_i/\mu$, $\tilde{q}_i=q_i/e$, $\tilde{\bf r}_i={\bf r}_i/(\sqrt{2}r_0)$, and $\tilde{r}_*=r_*/(\sqrt{2}r_0)$ be magnetic field, mass, charge, and position of particle $i$. The screening length and the Hamiltonian $\hat{\tilde{H}}=\hat{H}/E_0$ in logarithmic e.u. are as defined in Sec.\ref{sec:units}. We thus get:
\begin{eqnarray}
\hat{\tilde{H}} & = & -\sum_i \frac{1}{2\tilde{m}_i}\tilde{\nabla}_i^2+\sum_i \frac{\tilde{B}^2\tilde{r}_i^2}{8\tilde{m}_i}-\sum_{i>j} \tilde{q}_i\tilde{q}_j\ln\left(e^\gamma\tilde{r}_{ij}\right/2)\nonumber \\ & & {}+\sum_{i>j}\tilde{q}_i\tilde{q}_j\ln\left(\tilde{r}_*\right),
\end{eqnarray}
where we neglect the term $(i\hbar q_i/m_i) {\bf A}_i\cdot\nabla_i$ in Eq.\ref{eq:hamiltonian} that breaks time-reversal symmetry. The energy eigenvalue $\tilde{E}=E/E_0$ is therefore the sum of a function $f(\sigma,\tilde{B})$, where $\sigma=m_{\rm e}/m_{\rm h}$, and an additive constant $c(\tilde{r}_*)=\sum_{i>j}\tilde{q}_i \tilde{q}_j \ln \left(\tilde{r}_*\right)$. For X in the absence of an external magnetic field, $\tilde{E}^{\rm X}_{B=0}=0.41057747491(7)-\ln(\sqrt{2})-\ln(\tilde{r}_*)$ was calculated in Ref.\onlinecite{Mostaani_2017}.

For $\tilde{B}$ such that the magnetic confinement energy is larger than the log interaction, the interaction $-\sum_{i>j}\tilde{q}_i\tilde{q}_j\ln(e^\gamma \tilde{r}_{ij}/2)$ is negligible compared with the magnetic confinement energy of each particle. The dimensionless total energy is the sum of the zero-point energies of the individual particles in the quadratic potential plus the constant $c(\tilde{r}_*)$. Hence, at large $\tilde{B}\gg 1$:
\begin{align}
\tilde{E} & = \left( \frac{N_{\rm e}}{\tilde{m}_{\rm e}}+\frac{N_{\rm h}}{\tilde{m}_{\rm h}} \right)\frac{\tilde{B}}{2}+O(1)+c(\tilde{r}_*)\\ &
\approx\left(\frac{N_{\rm e}}{\tilde{m}_{\rm e}}+\frac{N_{\rm h}}{\tilde{m}_{\rm h}}\right)\frac{\tilde{B}}{2} +\tilde{E}_{\tilde{B}=0},\label{eq:CB-largeB}
\end{align}
since $\tilde{E}_{\tilde{B}=0}\sim c(\tilde{r}_*)$, when $\tilde{r}_*$ is large ($\tilde{r}_*\gg 1$).
\begin{figure}
\centerline{\includegraphics[width=90mm]{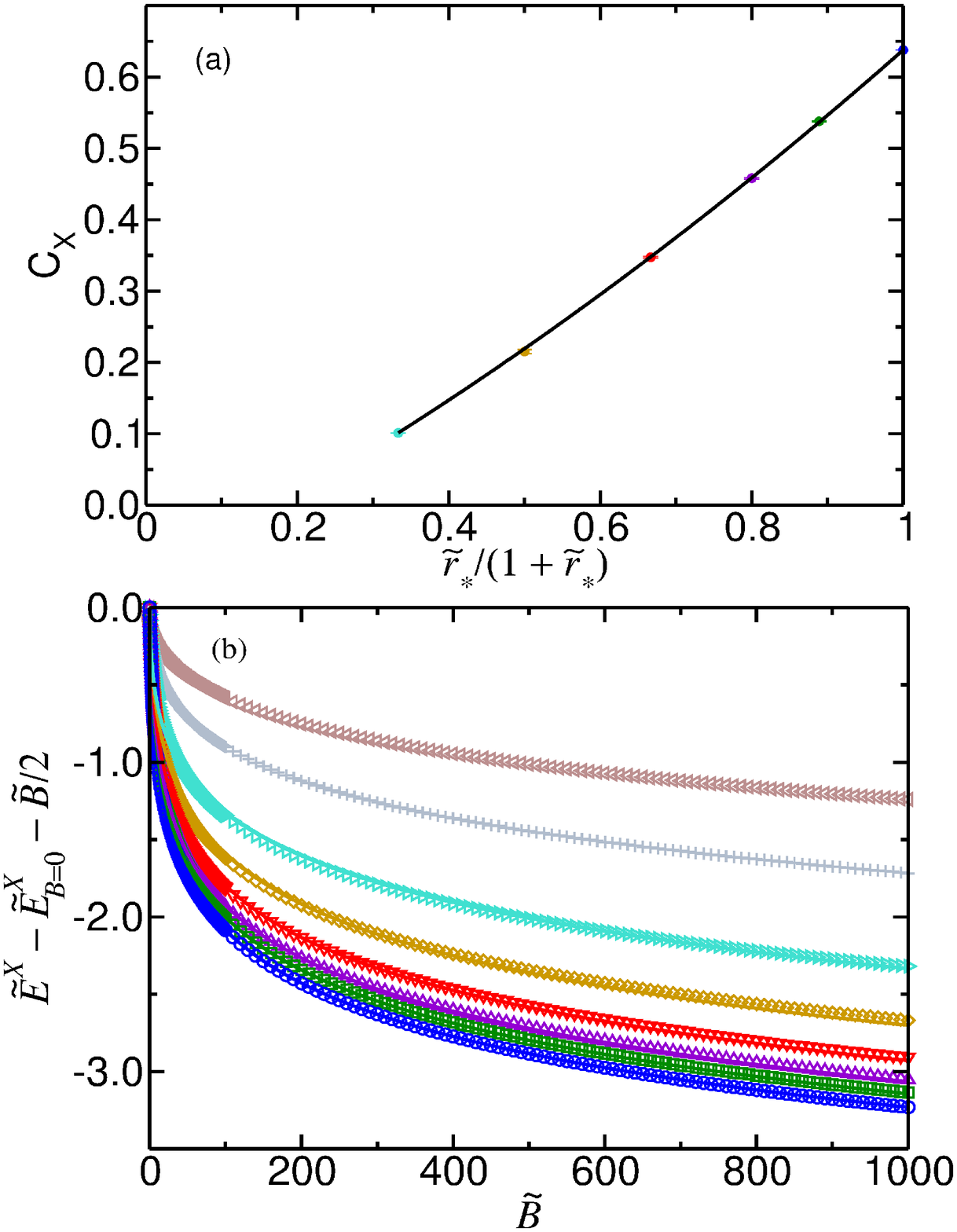}}
\centerline{\includegraphics[width=60mm]{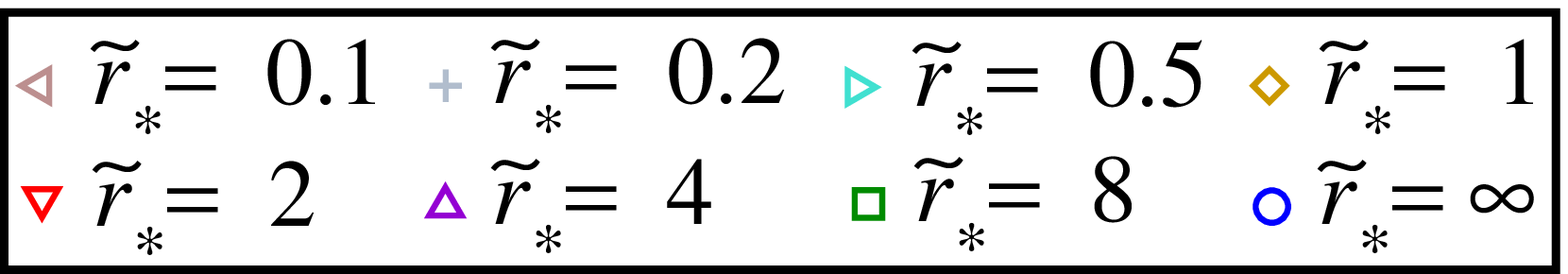}}
\caption{(a) Dependence of $C_{\rm X}$ on susceptibility. The markers show fitted $C_{\rm X}$. The line is a quadratic fit to the points as for Eq.\ref{eq:C_ex_fit}.(b) Shift in energy of X with equal $m_{\rm e}^*$ and $m_{\rm h}^*$ due to external magnetic field, after subtracting the large-$\tilde{B}$ behavior from Eq.\ref{eq:CB-largeB}, for several $\tilde{r}_*$. Markers indicate the finite-element method results, while lines show the fit of Eq.\ref{eq:CBfit}.\label{fig:CB}}
\end{figure}

For small $\tilde{B}\ll 1$, we use the CoM zero-point energy approximation, Eq.\ref{eq:com_zpe}, in which we assume the quadratic potential varies on the scale of the complex. Then:
\begin{equation}
\tilde{E}=\frac{\tilde{B}}{2}\sqrt{\frac{N_{\rm e}/\tilde{m}_{\rm
      e}+N_{\rm h}/\tilde{m}_{\rm h}}{N_{\rm e}\tilde{m}_{\rm
      e}+N_{\rm h}\tilde{m}_{\rm h}}} + \tilde{E}_{\tilde{B}=0}.
\end{equation}
The total energies for X with $m_{\rm e}=m_{\rm h}$ are calculated using the finite-element method (FEM) implemented in Mathematica\cite{Mathematica}. The results are converged by increasing the region size and decreasing the maximum cell size in order to achieve at least six digits of precision. This leads to errors comparable errors to QMC (see Sec.\ref{app:qmc}). Subtracting the large-$\tilde{B}$, Eq.\ref{eq:CB-largeB}, from the energy shift of X due to external magnetic fields, results in the logarithmic-like behavior in Fig.\ref{fig:CB}. This suggests the following formula for the energy shift of a generic charge-carrier complex due to external magnetic field:
\begin{align}
\tilde{E}-\tilde{E}_{B=0}=\frac{1}{2}&\left[\sqrt{\frac{N_{\rm e}/\tilde{m}_{\rm e}+N_{\rm h}/\tilde{m}_{\rm h}}{N_{\rm e}\tilde{m}_{\rm e}+N_{\rm h}\tilde{m}_{\rm h}}}-\right. \nonumber\\&\left. \left(\frac{N_{\rm e}}{\tilde{m}_{\rm e}}+\frac{N_{\rm h}}{\tilde{m}_{\rm h}}\right)\right]\ln\left(1+\tilde{B}+C^2\tilde{B}^2\right)\nonumber\\ &+\left(\frac{N_{\rm e}}{\tilde{m}_{\rm e}}+\frac{N_{\rm h}}{\tilde{m}_{\rm h}}\right)\frac{\tilde{B}}{2},
\label{eq:CBfit}
\end{align}
where $C=C(\tilde{r}_*,\sigma)$ is independent of $\tilde{r}_*$ in the $r\ll r_*$ limit in which the logarithmic interaction is valid.

We use the least-squares method to fit the FEM results for the BE shift of X with equal $m_{\rm e}^*$ and $m_{\rm h}^*$, Fig.\ref{fig:CB}, for several values of susceptibility, and extract the fitting parameter $C_{\rm X}$ for each $\tilde{r}_*$. We use a polynomial fit to get the dependence of $C_{\rm X}$ on susceptibility, Fig.\ref{fig:CB}:
\begin{equation}
C_{\rm X}=-0.1020(22)+0.546(9)x+0.194(6)x^2,
\label{eq:C_ex_fit}
\end{equation}
where $x=\tilde{r}_*/\left(1+\tilde{r}_*\right)$.  Since most 1L-TMDs have effective mass ratios close to 1, Table \ref{table:compare_quint}, we neglect the mass-ratio dependence of $C$.  Our fit is only valid for $\tilde{r}_*\gtrsim 0.5$. Although the fit from Eq.\ref{eq:CBfit} is derived for the logarithmic interaction, it does fit well our DMC results in Fig.\ref{fig:wse2} for the full Keldysh interaction for experimentally relevant values, as shown by the red curves in Fig.\ref{fig:wse2}.

\end{document}